\documentclass[onecolumn,sort&compress,numbers]{els-mrw} 

\usepackage{amsmath,amssymb,amsfonts,amsthm,makeidx,graphicx,subcaption}
\usepackage{txfonts}
\usepackage{helvet}

\usepackage{slashed}
\usepackage{gensymb}
\usepackage{bbold}
\usepackage{multirow}

\usepackage{bm}
\usepackage{bbm}
\usepackage{makecell}
\usepackage[dvipsnames]{xcolor}
\usepackage[normalem]{ulem}
\usepackage{longtable}
\usepackage{array}
\usepackage{float}
\usepackage{esvect}
\newcommand*{\Rom}[1]{\uppercase\expandafter{\romannumeral #1\relax}}

\newcommand{\beq}{\begin{equation}}
\newcommand{\eeq}{\end{equation}}
\newcommand{\bea}{\begin{eqnarray}}
\newcommand{\eea}{\end{eqnarray}}

\def\vec#1{{\bf #1}}

\renewcommand{\arraystretch}{1.2}
\newcommand{\abs}[1]{\lvert#1\rvert}

\begin{document}


\chapter{Baryon Form Factors}\label{chap1}

\author[1,2]{Yong-Hui Lin}%
\author[2,3]{Hans-Werner Hammer}%

\author[1,4,5]{Ulf-G. Mei{\ss}ner}%

\address[1]{\orgname{Universit\"at Bonn}, \orgdiv{Helmholtz Institut f\"ur Strahlen- und Kernphysik and Bethe Center
   for Theoretical Physics}, \orgaddress{53115 Bonn, Germany}}
\address[2]{\orgname{Technische Universit\"at Darmstadt}, \orgdiv{Department of Physics}, \orgaddress{64289 Darmstadt, Germany}}
\address[3]{\orgname{GSI Helmholtzzentrum f\"{u}r Schwerionenforschung GmbH}, \orgdiv{ExtreMe Matter Institute EMMI and Helmholtz Forschungsakademie Hessen f\"ur FAIR (HFHF)}, \orgaddress{64291 Darmstadt, Germany}}
\address[4]{\orgname{Forschungszentrum J\"ulich}, \orgdiv{Institute for Advanced Simulation (IAS-4)}, \orgaddress{52425 J\"ulich, Germany}}
\address[5]{\orgname{Beihang University}, \orgdiv{Peng Huanwu Collaborative Center for Research and Education},
\orgaddress{Beijing 100191, China}}

\articletag{Chapter Article tagline: update of previous edition, reprint.}

\maketitle

\begin{abstract}[Abstract]
We review the status of baryon form factors with a special focus on the nucleon electromagnetic form factors which are known best. First, we give an introduction into  the dispersive analyses and emphasize the role of unitarity and analyticity
in the construction of the isoscalar and isovector spectral functions.
Second, we present the state of the art in our understanding of nucleon form factors and radii including reliable uncertainty estimates from bootstrap and Bayesian methods.
Third, we discuss the physics of the time-like form factors
and point out further issues to be addressed in this framework. Finally, we review the status of hyperon form factors and comment on the pion cloud.
\end{abstract}

\begin{keywords}
 	nucleon structure\sep form factors\sep dispersion theory\sep nucleon radii\sep  hyperon structure
\end{keywords}

\section{Introduction}\label{intro}

Baryons are composite particles made of three valence quarks bound together by the strong force.  Understanding their structure and interactions is crucial for unraveling the complexities of quantum chromodynamics (QCD)~\cite{Wilson:1974sk,Wilczek:2012sb}, the fundamental theory describing strong interactions in terms of quarks and gluons~\cite{Gross:2022hyw}.
One of the key concepts for describing the internal structure of baryons and other 
strongly interacting particles (hadrons) are form factors. 
The form factors characterize how the response of composite objects to external probes differs from point particles.
They describe the distribution of properties like charge, current, and spin within hadrons. Form factors are typically studied through scattering experiments, where a lepton interacts with a baryon via the exchange of a photon or a weak gauge boson. 
A theoretical discussion of the structure 
of baryons from the perspective of chiral perturbation theory  and Schwinger-Dyson equations can, e.g., be found in Refs.~\cite{Kubis:2000aa} and \cite{Eichmann:2016yit}, respectively.

In this chapter, we focus primarily on the electromagnetic (EM) form factors of the nucleon from the viewpoint of dispersion theory
\cite{Lin:2021umz}.
The nucleon ($N$) form factors are best understood experimentally and theoretically. They play a fundamental role
since protons ($p$) and neutrons ($n$) essentially account for all the mass of everyday matter.
The EM form factors of the nucleon describe the
structure of the nucleon as seen by an electromagnetic probe.
As such, they provide a window on the strong interaction dynamics
in the nucleon from large to small distances. At small momentum transfers, they probe the large-distance properties of the nucleon like the charge and magnetic moment, while the quark substructure of the nucleon determines the behavior at large momentum transfer. For recent reviews on the form factors see, e.g. Refs.~\cite{Denig:2012by,Pacetti:2014jai,Punjabi:2015bba,Lin:2021umz}.
The form factors enter in the description of
a wide range of physical quantities ranging from the Lamb shift in atomic physics \cite{Pohl:2010zza,Beyer:2017gug,Fleurbaey:2018fih,Bezginov:2019mdi} over the strangeness content of the
nucleon~\cite{Armstrong:2012bi,Maas:2017snj}
to the EM structure and reactions of atomic nuclei
\cite{Bacca:2014tla,Phillips:2016mov,Krebs:2020pii}.

In contrast to the nucleon case, the experimental knowledge of the electromagnetic structure of other baryons and, 
in particular, of those with strangeness (the hyperons) is scarce due to absence of stable targets. 
So far, only the charge radius of the $\Sigma^-$, among all hyperons, 
has been measured to be $0.78(10)$~fm using a $\Sigma^-$ beam at a mean energy of 610~GeV~\cite{SELEX:2001fbx}.
A new experimental approach to access the charge radii of charged hyperons was recently proposed by Ref.~\cite{Lin:2023qnv}. 
This technique extracts the low-$t$ electromagnetic form factors in the unphysical region 
from the radiative Dalitz decay of charmonium, $\psi(2S) \to Y\bar{Y}e^+e^-$.
The main source of information on $\Lambda$ hyperon EM form factors in the time-like region
are measurements of the reaction $e^+ e^- \to \Lambda\bar{\Lambda}$ (see Ref.~\cite{Zhou:2022jwr} for a review). A recent improvement of the data base for this reaction is provided by cross sections in the 
center-of-mass energy region from 3.51 to 4.6 GeV by the BESIII Collaboration \cite{BESIII:2021ccp}, since previous measurements only covered the near-threshold region. 


\section{Basics}\label{sec:for}

\subsection{Definitions}\label{sec:for:subsec1}
The EM nucleon form factors and radii are determined by
the matrix element of the electromagnetic current operator
$j_\mu^{\rm EM}(x)$ in the nucleon. 
Denoting a nucleon state with four-momentum $p$ and spin $s$ as $|N(p,s)\rangle$,
this matrix element can be expressed as, 
\beq\label{eq:NME}
\langle N(p', s') | j_\mu^{\rm EM}(0) | N(p, s) \rangle = \bar{u}(p',s')
\left[ F_1 (t) \gamma_\mu +i\frac{F_2 (t)}{2 m} \sigma_{\mu\nu}
q^\nu \right] u(p,s)\,
\eeq
where $m$ is the nucleon mass 
and $t=(p'-p)^2$  the four-momentum transfer
squared. For space-like momentum transfer $t<0$, it is convenient to use the
variable $Q^2=-t>0$. The scalar functions $F_1(t)$ and $F_2(t)$ are the Dirac and 
Pauli form factors, respectively. Their normalization at $t=0$
is determined by the charges and anomalous magnetic moments of the nucleon,
\begin{equation}
\label{norm}
F_1^p(0) = 1\,, \quad  \; F_1^n(0) = 0\,,\quad  F_2^p(0) =  \kappa_p\,, \quad  F_2^n(0) = \kappa_n\, ,
\end{equation}
with $\kappa_p=1.793$ and $\kappa_n=-1.913$ given in units of the nuclear magneton, $\mu_N = e/(2m)$.
The  magnetic moments of the proton and the neutron are thus given by $\mu_{p} = 1 + \kappa_{p}$
and $\mu_n=\kappa_n$, respectively.

For the dispersion-theoretical analysis, it is convenient to work in the isospin basis and to 
decompose the form factors into isoscalar ($s$) and isovector ($v$) parts,
\begin{equation}
F_i^s = \frac{1}{2} (F_i^p + F_i^n) \, , \quad
F_i^v = \frac{1}{2} (F_i^p - F_i^n) \, , \quad
i = 1,2 \,. 
\end{equation}
The experimental data are usually given in terms of the Sachs form factors, 
\begin{equation}
G_{E}(t) = F_1(t) - \tau F_2(t) \, ,\quad G_{M}(t) = F_1(t) + F_2(t) \, , \quad\mbox{where}\quad \tau = -t/(4 m^2)\,,
\label{sachs}
\end{equation}
because the differential cross sections have a simple dependence on the Sachs form factors (see below). Their definition, Eq.~(\ref{sachs}), implies that $G_E=G_M$
at the nucleon-antinucleon threshold, $t = 4m^2$.
In the Breit frame, where the electromagnetic current transfers only three-momentum but no energy, the Sachs form factors $G_{E}$ and $G_{M}$ may be interpreted as the Fourier transforms of the charge and magnetization distributions, respectively. Note, however,  that it has been repeatedly pointed out that 
the identification of spatial density distributions with the Fourier transform of the corresponding form
factors in the Breit frame is problematic, see e.g. Refs.~\cite{Burkardt:2000za,Miller:2007uy,Jaffe:2020ebz}, and much work has been done  to interpret  the spatial density distributions of matrix elements of local operators~\cite{Miller:2010nz,Guo:2021aik,Panteleeva:2021iip,Lorce:2018egm,Freese:2021czn,Epelbaum:2022fjc,Panteleeva:2022khw}.

The nucleon root mean square radii (loosely called radii), $r \equiv \sqrt{\langle r^2 \rangle}$,
are defined via the expansion of the form factors around 
zero momentum transfer,
\beq
\label{def:r2}
\frac{F(t)}{F(0)}=1+t \frac{\langle r^2 \rangle}{6} +\ldots\,,
\eeq
where $F(t)$ is a generic form factor. In the case of the electric
and Dirac form factors of the neutron, $G_E^n$ and $F_1^n$, the
expansion starts with the term linear in $t$ and the 
normalization factor $F(0)$ is dropped. 
In the space-like momentum transfer region, $t<0$, the form factors are real valued quantities.
In the time-like region, $t>0$, the form factors are complex valued above the two-pion threshold at $t=4M_\pi^2$ and thus throughout the physical region of the process $e^+e^-\to \bar NN$.

\begin{figure}[t] 
\centerline{\includegraphics*[width=0.45\linewidth,angle=0]{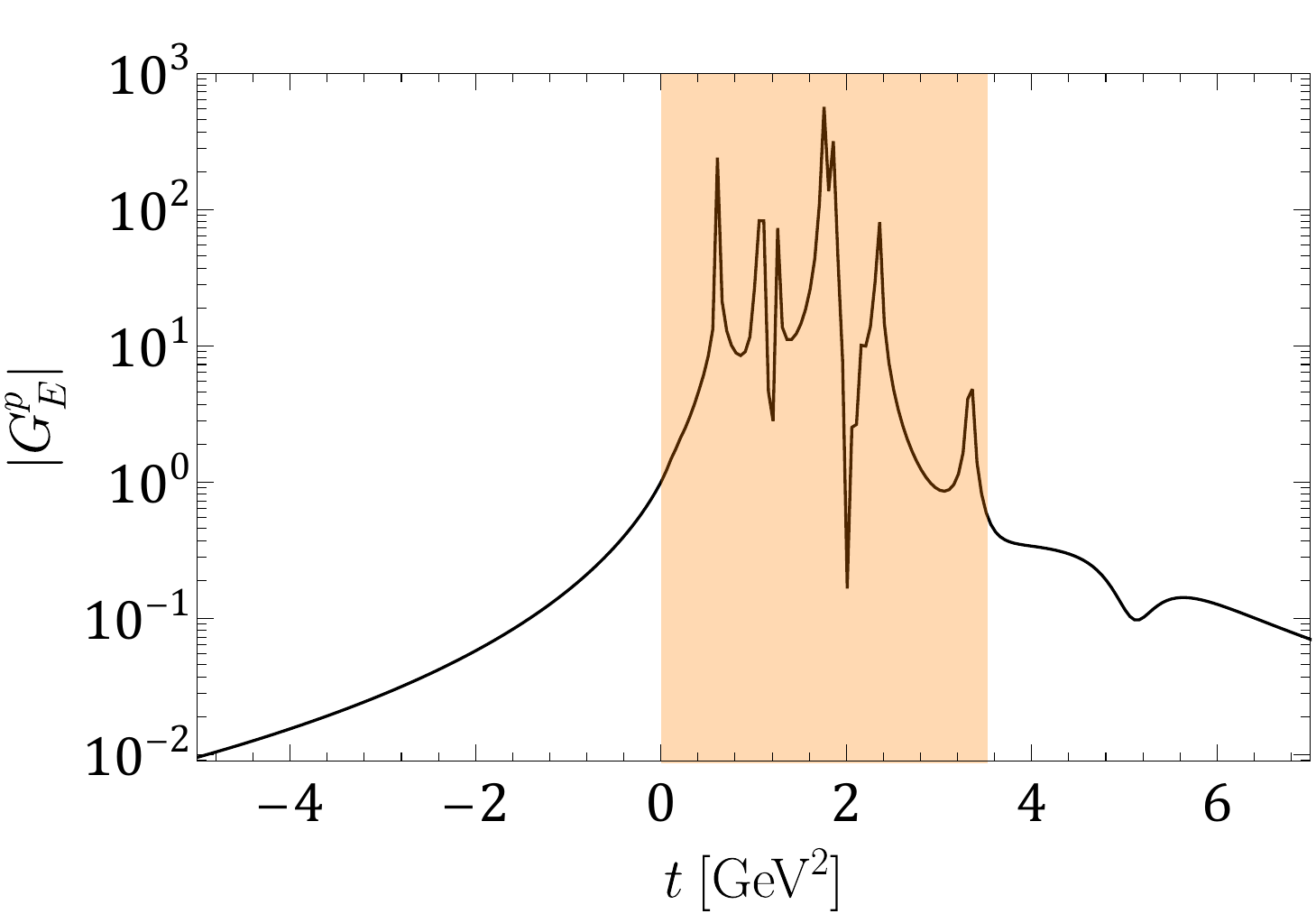}\quad\includegraphics*[width=0.45\linewidth,angle=0]{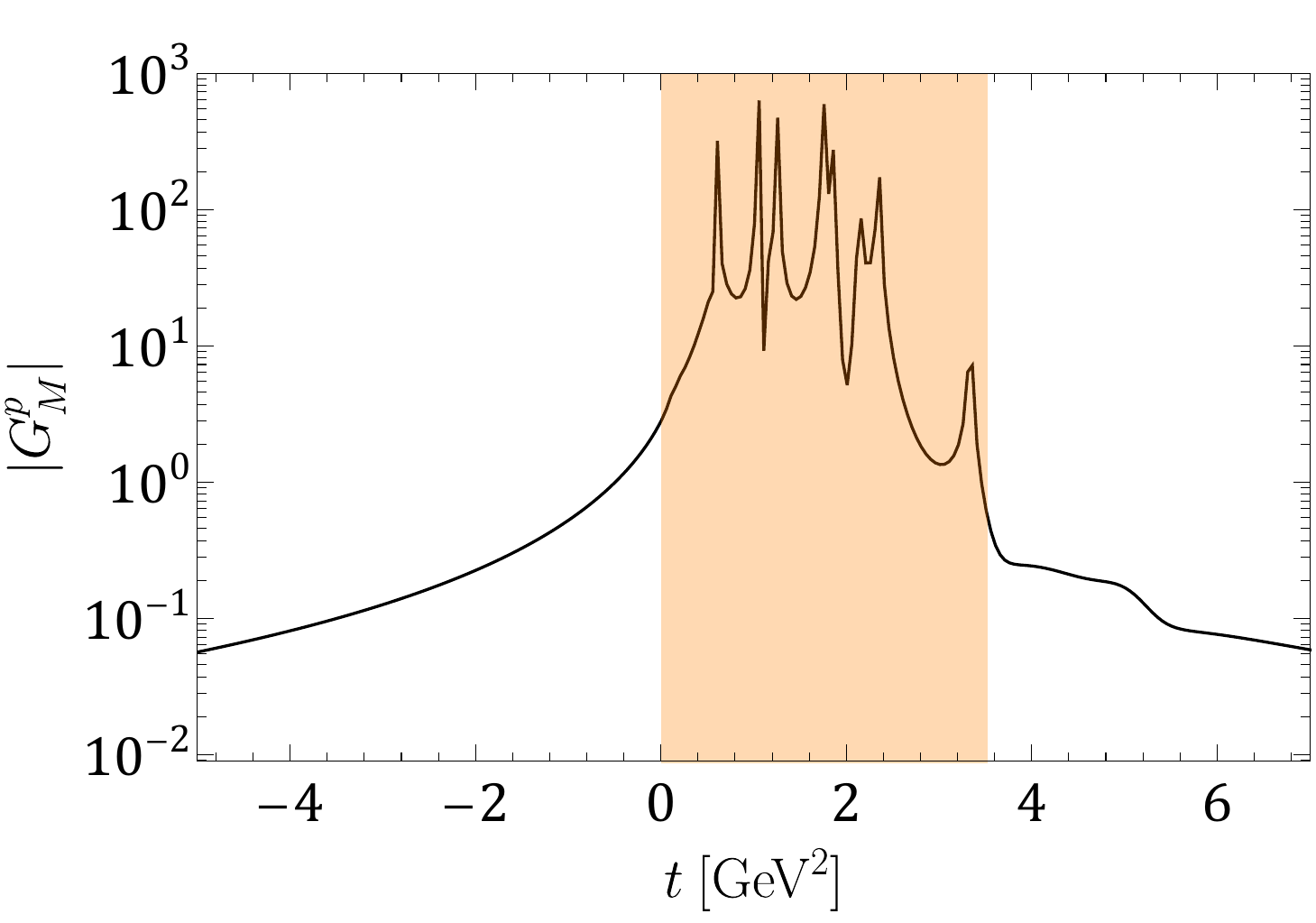}}
\caption{The moduli of $G_E^p$ (left panel) and $G_M^p$ (right-panel) for space- and time-like momentum transfers based on a recent dispersion-theoretical analysis of form factor data~\cite{Lin:2021xrc}.
The colored area between the two dashed lines at $t=0$ and $t=4m^2$ is the unphysical region where the form factor cannot be observed.}
\label{fig:FFgen}
\vspace{-3mm}
\end{figure} 
In Fig.~\ref{fig:FFgen}, we sketch 
$G_E^p(t)$ and of $G_M^p(t)$ as examples. More precisely, the moduli of these form factors is depicted. For the
space-like region, the physical threshold is located at $t=0$, whereas the corresponding
threshold in the time-like region is $t=4m^2$. In between these two thresholds,
the various vector meson poles (plus continua) build up the spectral function
to be discussed in detail below. This region cannot be observed. 
We note that for the form factors in the time-like
region, an additional complication arises due to the strong near-threshold nucleon-antinucleon
interactions. 

\subsection{Experimental Observables}
We start by discussing how the nucleon EM form factors can be accessed in experiment.

\subsubsection{Space-like region}

The space-like form factors (FFs) can be measured in elastic
electron scattering. We consider for definiteness electron-proton ($ep$) scattering,
\beq
e\,(p_1) + p\,(p_2) \to e\,(p_3) + p\,(p_4)~,
\eeq
where the four-momenta $p_i$ are subject to the constraint $p_1+p_2 = p_3+p_4$. 
At first order in the EM fine-structure constant $\alpha$, i.e. in the one-photon exchange approximation,
the differential cross section can be expressed through the Sachs FFs as
\begin{equation}\label{eq:xs_ros}
\frac{d\sigma}{d\Omega} = \left( \frac{d\sigma}
{d\Omega}\right)_{\rm Mott} \frac{1}{\epsilon (1+\tau)}
{\left[\tau G_{M}^{2}(Q^{2}) + {\epsilon} G_{E}^{2}(Q^{2})\right]}   = \left( \frac{d\sigma}
{d\Omega}\right)_{\rm Mott} \frac{\sigma_R}{\epsilon (1+\tau)}
\, ,
\end{equation}
where
$\epsilon = [1+2(1+\tau)\tan^{2} (\theta/2)]^{-1}\,$ with $\:0\leq \epsilon \leq 1$
is the virtual photon polarization,
$\theta$ is the electron scattering angle in the laboratory frame, and
$({d\sigma}/{d\Omega})_{\rm Mott}$ is the Mott cross section. The latter corresponds to scattering off
a point-like nucleon,
\begin{equation}\label{eq:xs_mott}
\left( \frac{d\sigma}{d\Omega}\right)_{\rm Mott} =
\frac{\alpha^2 \cos^2(\theta/2)}{4E_1^2\sin^4(\theta/2)} \frac{E_3}{E_1}~,
\end{equation}
where $E_1\ (E_3)$ is the energy of the incoming (outgoing) electron, related
via $1/E_3= 1/E_1 +(2/m) \sin^2(\theta/2)$.
Two quantities out of the energies, momenta and angles suffice to determine this cross
section and are related for such an elastic process. Specifically, in the laboratory frame
with the initial nucleon at rest and neglecting the electron mass, we can write the four-momentum transfer
\begin{align}
Q^2 \approx 4E_1E_3\sin^2\left({\theta}/{2}\right)~.
\end{align}
In experiment, the differential cross section is usually given for a fixed total energy
as a function of the scattering angle, so that a small scattering angle corresponds to
a small momentum transfer. This is exactly the reason why a precise determination of the
EM radii is so difficult.  At large momentum transfer, the contribution from the magnetic form factor
dominates the cross section.
The contribution from the electric and the magnetic form factor
can be read off form the reduced cross section $\sigma_R$ defined in Eq.~\eqref{eq:xs_ros}.
The reduced cross section $\sigma_R$ depends linearly on $\epsilon$ for a given $Q^2$, with slope $G_E^2(Q^2)$ and
intercept $\tau G_M^2(Q^2)$. This is called the Rosenbluth separation~\cite{Rosenbluth:1950yq}. 
Two-photon corrections to this cross section need to be accounted when analyzing the real experimental data of the elastic electron-proton scattering, 
see Refs.~\cite{Arrington:2011dn,Lorenz:2014yda,Afanasev:2017gsk,Ahmed:2020uso,Lin:2021umz} for more details.
Also, to investigate the neutron form factors,
one measures electron scattering of a light nucleus like deuterium or $^3$He. This requires,
however, an accurate few-body calculation to disentangle the neutron FF contribution from the
nuclear scattering cross section, as discussed briefly in Ref.~\cite{Lin:2021umz}.

In early $ep$ scattering experiments,  it was found that the form factors could be
well approximated by the dipole form, $G_{\rm dip}(Q^2)$,
\begin{equation}
G_E^p(Q^2) \simeq \frac{G_M^p(Q^2)}{\mu_p} \simeq  \frac{G_M^n(Q^2)}{\mu_n} \simeq G_{\rm dip}(Q^2)=\left(1+ {Q^2}/{M_{dip}^2}\right)^{-2}~,
\end{equation}
with $M_{dip}^2=0.71~{\rm GeV}^2$ the dipole mass. Moreover, $G_E^n(Q^2)=0$ in this approximation.
Employing these dipole FFs in the integrated cross section Eq.~\eqref{eq:xs_ros} defines the
so-called dipole cross section, $\sigma_{\rm dip}$. Often, the form factors or the measured
cross sections are given relative to $G_{\rm dip}(Q^2)$ and $\sigma_{\rm dip}$, respectively.

A method to directly measure the form factor ratio $G_E/G_M$ in polarized electron scattering
off the proton, $\vv{e}p\to\vv{e}p$ (or similarly for scattering off
the deuteron or $^3$He), was
proposed in Refs.~\cite{Akhiezer:1968ek,Arnold:1980zj}. A simultaneous measurement of the two recoil polarizations (longitudinal, $P_l$, and transverse, $P_t$) allows one to measure
directly the ratio
\begin{equation}
R_p \equiv \mu_p \frac{G_E^p}{G_M^p} = -\mu_p \sqrt{\frac{\tau(1+\epsilon)}{2\epsilon}}\frac{P_t}{P_l}~.
\end{equation}  
While this only determines the form factor ratio (and not the individual FFs), many
systematic uncertainties cancel out and make this observable an important benchmark
for any theoretical form factor calculation.

\subsubsection{Time-like region}
We now turn to the  determination of the form factors in the time-like region. Here one needs a reaction in which a nucleon-antinucleon pair is produced electromagnetically. 
The FFs can, e.g.,  be extracted from the cross section data $e^+e^- \leftrightarrow \bar{p}p$
and  $e^+e^-\to \bar{n}n$ for the proton and the neutron, respectively.
As only very few differential cross section data exist
in the time-like region, a separation of $G_E$ and $G_M$ is often not possible. One either makes an assumption like $G_E = G_M$ in the analysis of the data
or one extracts the effective form factor $|G_{\rm eff}|$, discussed below.
For a review on  the nucleon EM form factors in the time-like region, see Ref.~\cite{Denig:2012by}. 
Let us consider the process
\beq
e^+\,(p_1)+e^-\,(p_2)\to
p\, (p_3)+\bar{p}\,(p_4)~,
\eeq
in more detail. In the center-of-mass (CM) frame, we have $p_{1,2} = (E,\pm k_e)$ and $p_{3,4} = (E,\pm k_p)$.
The photon momentum $q$ then determines the center-of-mass energy by $q^2 = (p_1+p_2)^2=t
= E_{\rm CM}^2 = (2E)^2$. Time-like $q$ implies positive $q^2$ in our metric. The three-momenta $k_e$ and $k_p$ enter in the phase-space factor $\beta = k_p/k_e$,  which in the limit 
of neglecting the electron mass yields
\beq
\beta \approx k_p/E = \sqrt{1 - 4m_p^2/q^2}~,
\eeq
the velocity of the proton, and $m_p$ is the proton mass. Denoting the emission
angle of the proton by $\theta$, the differential cross section in the one-photon-exchange approximation is
\beq
\frac{d\sigma}{d\Omega}
= \frac{\alpha^2\beta}{4 q^2}C(q^2)\biggl[(1+\cos^2\theta)|G_M(q^2)|^2 + \frac{4m_p^2}{q^2}\sin^2\theta|G_E(q^2)|^2\biggr]~,
\eeq
where $C(q^2)$ is the Sommerfeld-Gamow
factor that accounts for the Coulomb interaction between the final-state particles
\begin{align}
 C(q^2)=\frac{y}{1-e^{-y}},\hspace{8pt} y=\frac{\pi\alpha m_p}{k_p}.
\end{align}
Integrating over the full angular distribution gives the total cross section
\begin{align}
\sigma_{e^+e^- \rightarrow p\bar{p}}(q^2) &= \frac{4\pi\alpha^2\beta}{3q^2}C(q^2)
\left[|G_M(q^2)|^2+\frac{2m_p^2}{q^2}|G_E(q^2)|^2\right]\equiv \frac{4\pi\alpha^2\beta}{3q^2}C(q^2)\left(1+\frac{2m_p^2}{q^2}\right)|G_{\rm eff}^p(q^2)|^2.
\end{align}
which defines the effective form factor
$G_{\rm eff}$.
For neutrons, the formulas are equivalent except for the Sommerfeld-Gamow factor which is
not present in that case. Beyond the Coulomb final-state interactions, higher order QED
corrections are usually neglected.  For the time-reversed process,
the phase space factor is inverted:
\beq
\sigma(e^+e^-\to p\bar{p}) = \beta^2\, \sigma( p\bar{p}\to e^+e^-)~.
\eeq
Polarization effects in $e^+e^-\to N\bar{N}$ are studied e.g. in Ref.~\cite{Dubnickova:1992ii}.

\subsection{Spectral Decomposition and Dispersion Relations}

Dispersive methods provide a powerful tool to extract the
form factors from experimental data.
Dispersion relations (DRs) are based on unitarity and general analyticity properties of the form factors. Analog to the Kramers-Kronig relations from classical electrodynamics \cite{Jackson:1998nia}, they relate the real and imaginary parts of the form factors.

The imaginary part ${\rm Im}\, F$ of a form factor $F$ 
can be obtained from a spectral decomposition \cite{Chew:1958zjr,Federbush:1958zz}. 
For this purpose, we consider the  electromagnetic 
current matrix element in the time-like region ($t>0$), which is 
related to the space-like region ($t<0$) via crossing symmetry.
This matrix element is given by
\begin{equation}
\label{eqJ}
J_\mu = \langle N(p_3)\, \bar{N}(p_4) | j_\mu^{\rm EM}(0) | 0 \rangle = \bar{u}(p_3) \left[ F_1 (t) \gamma_\mu +i\frac{F_2 (t)}{2 m} \sigma_{\mu\nu}
(p_3+p_4)^\nu \right] v(p_4)\,,
\end{equation}
where $p_3$ and $p_4$ are the momenta of the nucleon and antinucleon created by the current $j_\mu^{\rm EM}$, respectively. For ease of notation, we have suppressed the corresponding spin indices.
The four-momentum transfer squared
in the time-like region is $t=(p_3+p_4)^2$. 

\begin{figure}[t] 
\centerline{\includegraphics*[width=0.4\linewidth,angle=0]{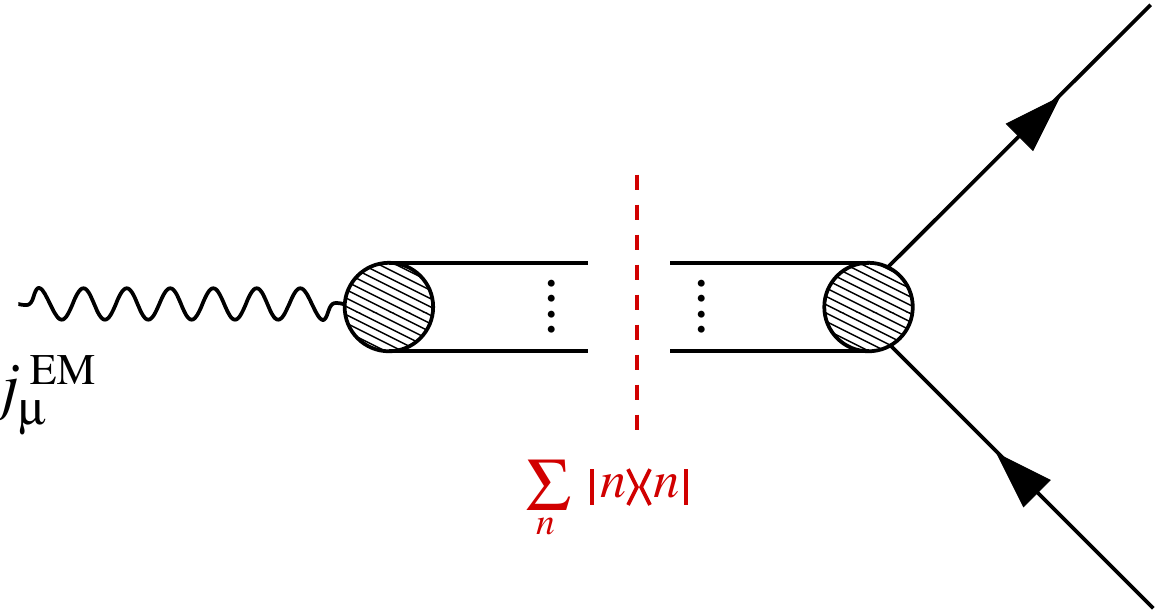}}
\caption{Spectral decomposition of the matrix element of the electromagnetic current $j_\mu^{\rm EM}$ in the nucleon. The intermediate states are denoted $|n\rangle$.}
\label{fig:specdeco}
\vspace{-3mm}
\end{figure} 
Using the LSZ reduction formalism, the imaginary part
of the form factors is obtained by inserting a complete set of
intermediate states \cite{Chew:1958zjr,Federbush:1958zz}
as illustrated in Fig.~\ref{fig:specdeco},
\begin{equation}
\label{spectro}
{\rm Im}\,J_\mu = \frac{\pi}{Z}(2\pi)^{3/2}{\cal N}\,\sum_n
\langle N(p_3) | \bar{J}_N (0) | n \rangle
\langle n | j_\mu^{\rm EM} (0) | 0 \rangle \,v(p_4)
\,\delta^{(4)}(p_3+p_4-p_n)\,,
\end{equation}
where ${\cal N}$ is a nucleon spinor normalization factor, $Z$ is
the nucleon wave function renormalization, and $\bar{J}_N (x) =
J_N^\dagger(x) \gamma_0^{}$ with $J_N(x)$ a nucleon source.
It relates the spectral function to on-shell matrix elements of other
processes, as detailed below.

The intermediate states $|n\rangle$ are asymptotic (observable) states of
total four-momentum $p_n$. They carry the {\em same} quantum numbers as
the current $j^{\rm EM}_\mu$:
\begin{eqnarray}
I^G(J^{PC}) &=& 0^-(1^{--})\, \qquad \mbox{for the isoscalar component}~,\nonumber\\
I^G(J^{PC}) &=& 1^+(1^{--})\, \qquad \mbox{for the isovector component}~,
\end{eqnarray}
of the current $j^{\rm EM}_\mu$. Here, $I$ and $J$ denote the isospin $I=0,1$
and the angular momentum $J=1$ of the photon, whereas $G$, $P$ and $C$ give
the $G$-parity, parity and charge conjugation quantum number, respectively.
Furthermore, these currents  have zero net baryon number. Because of $G$-parity, states
with an odd number of pions only contribute to the iso\-scalar
part, while states with an even number contribute to the 
isovector part.
For the isoscalar part  the lowest mass states are:
\begin{equation}
3\pi, 5\pi, \ldots, 
 K\bar{K}, K\bar{K}\pi, \ldots\ ,
\end{equation}
and for the isovector part they are:
\begin{equation}
2\pi, 4\pi, \ldots  K\bar{K}, \ldots\ . 
\end{equation}
Associated with each intermediate state is a
cut starting at the corresponding threshold in $t$ and running to
infinity. As a consequence,
the spectral function ${\rm Im}\, F(t)$ is different from zero along the
cut from $t_0$ to $\infty$, with $t_0 = 4 \, (9) \, M_\pi^2$ for the
isovector (isoscalar) case.

The spectral functions are the central quantities in the 
dispersion-theoretical approach. Using Eqs.~\eqref{eqJ} and \eqref{spectro}, they
can in principle be obtained from experimental data. 
In practice, this program can only be carried out for 
the lightest two-particle intermediate states. The full structure of the spectral functions for the
EM form factors of the nucleon is discussed in more detail below.
 
\begin{figure}[t] 
\centerline{\includegraphics*[width=0.4\linewidth,angle=0]{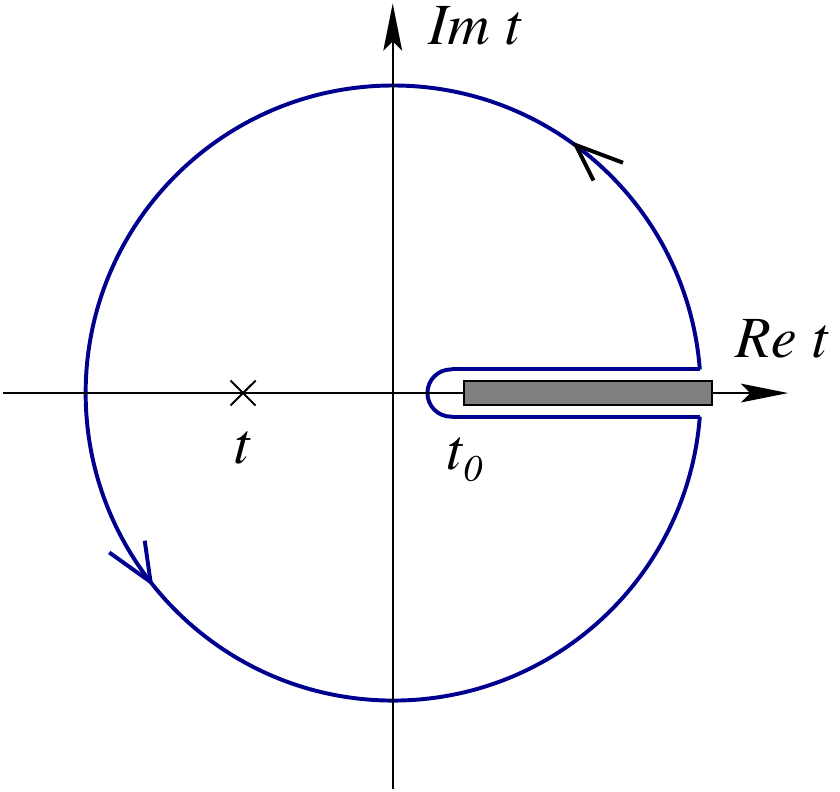}}
\caption{Analytic structure of a typical form factor in the complex plane. The start of the lowest continuum cut is indicated by $t_0$. The integration contour for the application of Cauchy's theorem to calculate the form factor $F(t)$ is shown in blue.}
\label{fig:disprel}
\vspace{-3mm}
\end{figure}

The typical singularity structure of a form factor $F$
resulting from the spectral decomposition, Eq.~(\ref{spectro}), is shown in Fig.~\ref{fig:disprel}.
We use Cauchy's theorem for $F$ with the integration contour indicated in blue to obtain an unsubtracted dispersion relation of the form
\begin{equation}
F(t) = \frac{1}{\pi} \, \int_{t_0}^\infty \frac{{\rm Im}\, 
F(t')}{t'-t-i\epsilon}\, dt'\, ,
\label{emff:disp} 
\end{equation}
where $t_0$ is the threshold of the lowest cut of $F(t)$
and the $i\epsilon$ defines the integral for values of $t$ on the cut.   
The convergence of an unsubtracted dispersion relation
for the form factors has been assumed. For proofs of such a representation
in perturbation theory, see Ref.~\cite{drell1961electromagnetic} (and references therein).
One could also use 
a once-subtrac\-ted dispersion relation, since the normalization of the
form factors at $t=0$ is known. However, in what follows, we will only
employ the unsubtracted form given in Eq.~\eqref{emff:disp}.  For the parametrization of the 
spectral functions used below, the unsubtracted dispersion relations converge by construction.
Consequently, by  Eq.~(\ref{emff:disp}) the electromagnetic structure
of the nucleon can be related to its absorptive behavior. 

\section{Spectral Function and Constraints}
We continue to discuss the structure of the spectral function
and constraints from theory and other processes.
\subsection{Structure of the spectral functions}
The longest-range and therefore at low momentum transfer most 
important continuum contribution comes from the $2\pi$ intermediate state
which contributes to the isovector form factors~\cite{Hohler:1974eq}.
A novel and very precise calculation of this contribution 
has recently been performed in Ref.~\cite{Hoferichter:2016duk} including
the state-of-the-art pion-nucleon scattering amplitudes from dispersion theory.
The resulting spectral functions 
exhibit the $\rho$-resonance at $\sqrt{t}=0.77\,$GeV as well as an enhancement on the left shoulder of the resonance. This confirms that the $\rho$ is
generated by unitarity~\cite{Frazer:1959gy} and no explicit $\rho$-meson is
required in the isovector spectral function.  The enhancement on the left shoulder
of the $\rho$ can be traced back
to the fact that the partial wave amplitudes $f_\pm^1(t)$ have a singularity on the
second Riemann sheet \cite{Hoehler:1983} (originating from the projection of the nucleon
pole terms in the invariant pion-nucleon scattering amplitudes) located at
\beq
t_c = 4M_\pi^2 - \frac{M_\pi^4}{m^2} \approx 3.98\,M_\pi^2~,
\eeq
very close to the physical threshold at $t_0 = 4M_\pi^2$. 

Based on the DR, Eq.~\eqref{emff:disp}, it is  straightforward to derive
sum rules  for the normalizations and radii of the isovector form factors.
These were first considered in Ref.~\cite{Hohler:1974eq} for the various
nucleon radii, see also~\cite{Hoferichter:2016duk},
\begin{align}\label{eq:SR1}
\frac{1}{2}(r^v_E)^2&=\frac{6}{\pi}\int\limits_{4M_\pi^2}^\infty  dt \frac{{\rm Im}~G_E^v(t)}{t^2}
=\frac{1}{2}\Big[(r^p_E)^2- (r^n_E)^2\Big]~,\notag\\
\mu^v(r^v_M)^2&=\frac{6}{\pi}\int\limits_{4M_\pi^2}^\infty dt \frac{{\rm Im}~G_M^v(t)}{t^2}
 =\frac{1}{2}\Big[(1+\kappa_p)(r^p_M)^2-\kappa_n(r^n_M)^2\Big]~,
\end{align}
where $\mu^v=(1+\kappa_p-\kappa_n)/2\simeq 2.353$ is the isovector magnetic
moment of the nucleon.
Note that the sum rules for the radii remain unchanged if a once-subtracted dispersion relation
is used instead of the unsubtracted one. Cutting the integrals off at $\Lambda = 2m$, one finds
\bea
\label{eq:SR}
\frac{1}{2}(r^v_E)^2 &=& 0.405(36)~{\rm fm}^2 \qquad
\mathrm{and}\qquad
\mu^v (r^v_M)^2 = 1.81(11)~{\rm fm}^2~.
\eea
It is remarkable that using a simple $\rho$-exchange with a Breit-Wigner or a Gounaris-Sakurai form \cite{Gounaris:1968mw}, the corresponding isovector radii would be
underestimated by about 40\%.
Thus, any dispersive analysis that does not
include the full two-pion continuum but only the $\rho$-resonance in the isovector spectral
function below 1~GeV will simply miss this important piece of physics.

The lowest isoscalar continuum is given by three-pion exchange. An analysis based on unitarity alone
of this contribution does not exist, but it has been shown in chiral perturbation theory
at leading \cite{Bernard:1996cc}  and subleading \cite{Kaiser:2019irl} orders, that there is 
no enhancement on the left wing of the $\omega$ resonance. 
Thus, in contrast to the $\rho$, the inclusion of the $\omega$ as a vector meson pole is justified.  The first significant continuum contribution to the isoscalar spectral function is due to $K\bar{K}$ and $\rho\pi$ intermediate states. 
The contributions from the $K\bar{K}$ \cite{Hammer:1998rz,Hammer:1999uf}
and $\rho\pi$ \cite{Meissner:1997qt}
were first included in the dispersive analysis of the EM form factors in Ref.~\cite{Belushkin:2006qa}. For recent
work on the isoscalar spectral functions in baryon chiral perturbation theory
with explicit vector mesons, that strengthens the findings of these earlier
works, see  Ref.~\cite{Unal:2019eum}. 
A more detailed discussion of these continua can be found in Ref.~\cite{Lin:2021umz}.

\begin{figure}[tb] 
\centering
\includegraphics*[width=0.7\textwidth,angle=0]{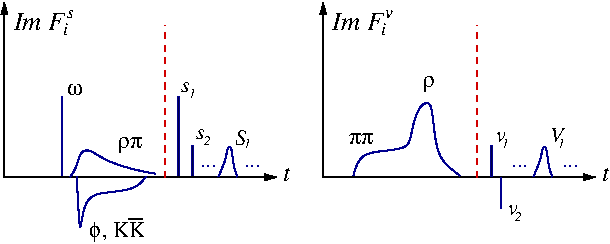}
\caption{Illustration of the isoscalar (left) and isovector (right) spectral function
  in terms of continua and (effective) vector meson poles. The vertical dashed
  line separates the well-constrained low-mass region from
  the high-mass region which is parameterized by narrow and broad effective poles, indicated by lower and upper case letters respectively.}
\label{fig:cartoon}
\vspace{-3mm}
\end{figure}

The remaining contributions to the spectral function
can be parameterized by vector meson poles. On the one hand, 
the lower mass poles can be identified with physical vector 
mesons such as the $\omega$ and the $\phi$.
The higher mass poles on the other hand, are simply an effective way
to parameterize higher mass strength in the spectral function.
These effective poles at higher momentum transfers appear 
in both the isoscalar and isovector channels. 
Note that the contributions from the continua and
the poles are sometimes strongly intertwined, e.g. the $\rho$-meson pole is indeed
generated as part of the $2\pi$-continuum, as known since
long~\cite{Frazer:1959gy,Frazer:1960zza,Frazer:1960zzb}.
It should also be noted that we are dealing with an ill-posed problem \cite{Ciulli:1975sm,SabbaStefanescu:1978hvt}. This implies that increasing the number of poles will
from some point on not improve the description of the data.
Therefore, the strategy has always been to use as few poles as possible. 
We come back to this issue in Sec.~\ref{sec:error}. 
The (isoscalar and isovector) spectral functions thus take the from
\begin{eqnarray}
{\rm Im }\,F_i^{s} (t) &=& {\rm Im }\,F_i^{(s,K\bar{K})} (t)
+ {\rm Im }\,F_i^{(s,\rho\pi)} (t)
+ \sum_{V=\omega,\phi,s_1,...} \pi a_i^{V}
\delta (M^2_{V}-t) 
+ \sum_{V=S_1,...}  {\rm Im }\,F_i^{(s,V)} (t) \, ,
\label{emff:s}\\
{\rm Im }\,F_i^{v} (t) &=& {\rm Im }\,F_i^{(v,2\pi)} (t)
+ \sum_{V= v_1,...} \pi a_i^{V} \delta (M^2_{V}-t)
+ \sum_{V=V_1,...}  {\rm Im }\,F_i^{(v,V)} (t)
\,, 
\label{emff:v}
\end{eqnarray}
where $i = 1,2\,$ and the broad effective poles are parameterized by a Breit-Wigner form.
The masses of all effective poles and the widths of the broad poles are fitted to the data. Moreover, all vector meson
coupling constants are fitted. 
A cartoon of these spectral functions is shown in Fig.~\ref{fig:cartoon}.  
The vertical dashed
line separates the phenomenologically well-constrained low-mass region from
the effective vector meson poles at higher masses.
Here, we allow a variable nonzero width for certain effective poles to mimic the imaginary part 
of the form factors in the higher-$t$ time-like region, as was done in Ref.~\cite{Belushkin:2006qa}.

\subsection{Constraints}
\label{sec:con}
The number of parameters in the spectral function (i.e. the various
meson couplings $a_i^V$ ($i=1,2$), the masses and widths of the effective poles)
is reduced by enforcing various constraints.

The first set of constraints concerns the low-$t$ behavior of the
form factors. We enforce the correct normalization of the form factors
as given in Eq.~(\ref{norm}). The nucleon radii, however, are not 
included as a constraint. The exception to this is the squared neutron
charge radius, which in some dispersive fits has been constrained
to the value from low-energy neutron-atom scattering
experiments \cite{Kopecky:1995zz,Kopecky:1997rw}. In the new fits discussed
later, we implement this constraint using the high-precision determination
of the neutron charge radius squared based on a chiral effective field theory
analysis of electron-deuteron scattering~\cite{Filin:2019eoe,Filin:2020tcs},
\beq\label{eq:r2En}
\langle r_n^2\rangle = -0.105^{+0.005}_{-0.006}~{\rm fm}^2~.
\eeq

Another set of constraints arises at large momentum transfers.
Perturbative QCD (pQCD) constrains the behavior of the nucleon
electromagnetic form factors for large momentum transfer.
Brodsky and Lepage \cite{Lepage:1980fj} worked out the behavior  for $Q^2 \to \infty$,
\begin{equation}
F_i (t) \to \frac{1}{Q^{2(i+1)}} \, 
\left[ \ln\left(\frac{Q^2}{ \Lambda_{\rm QCD}^2}\right)
\right]^{-\gamma} \, , \quad i = 1,2 \, ,
\label{emff:fasy1}
\end{equation}
where $\gamma = 2 + 4/(3\beta)$ is an anomalous dimension and $\beta = 11 - 2N_f/3$ the leading order QCD $\beta$-function.
The anomalous dimension $\gamma\approx 2$ depends weakly on the number of
flavors, $N_f$ \cite{Lepage:1980fj}.
The power behavior of the form factors at large $Q^2$ can be easily 
understood from perturbative gluon exchange. In order to distribute the 
momentum transfer from the virtual photon
to all three quarks in the nucleon, at least two massless
gluons have to be exchanged. Since each of the gluons has a propagator 
$\sim 1/Q^2$, the form factor has to fall off as $1/Q^4$. In the case
of $F_2$, there is additional suppression by $1/Q^2$ since a quark spin 
has to be flipped. The analytic continuation of the logarithm in Eq.~\eqref{emff:fasy1}
to time-like momentum transfers $-Q^2\equiv t>0$ yields an additional term,
$\ln(-t/\Lambda^2) =
\ln(t/\Lambda^2) - i\pi$ for $t> \Lambda^2$. The Phragmen-Lindeloef
theorem~\cite{Hoehler:1983} therefore stipulates that  the imaginary part has to vanish
in the asymptotic limit. Taking these facts into account, the proton effective
FF can be described for large time-like momentum transfer $t$
by~\cite{Bianconi:2015owa}
\begin{align}
 |G_{\rm eff}^p(t)| = \frac{A}{t^2(\ln^2(t/\Lambda^2)+\pi^2)},\label{eq:pqcd}
\end{align}
with the parameters from a fit to data prior to the 2013 measurement by the
BaBar collaboration~\cite{BaBar:2013ves}, given as $A=72\,$GeV$^{-4}$ and $\Lambda=0.52\,$GeV.

Enforcing the power law behavior of the form factors in the dispersion relation, Eq.~(\ref{emff:disp}), leads to superconvergence relations of the form
\begin{equation}
\label{eq:scr}
\int_{t_0}^\infty {\rm Im}\, F_i (t) \;t^n dt =0\, , \quad i = 1,2 \, ,
\end{equation}
with $n=0$ for $F_{1}$ and $n=0,1$ for $F_{2}$. These will be employed
in the current analysis. In earlier DR analyses, modifications of the
superconvergence relations
were used including e.g. some higher order corrections. These should,
however, be abandoned as the data are simply not sensitive to such corrections.
We note that these  superconvergence relations have already been used in
Ref.~\cite{Hohler:1976ax}, i.e. before the pQCD analysis~\cite{Lepage:1980fj}.

Consequently, the number of effective poles is determined
by the stability criterion mentioned before, that is,
we take the minimum number of poles necessary to fit the data.
The number of free parameters is then strongly reduced by the 
various constraints (unitarity, normalizations, superconvergence
relations). These constraints can be implemented as what is called
``hard constraints'' or ``soft constraints'', respectively. In the
former case, one solves a system of algebraic equations relating the
various parameters (couplings, masses), thus reducing the number of free
parameters in the fit (for an explicit representation,
see e.g.~\cite{Mergell:1995bf}). In the latter case, the $\chi^2$ is
augmented by a Lagrange multiplier enforcing the corresponding constraints, see
Sec.~\ref{sec:error}. Both options are viable and have been used.

It is straightforward to enumerate the number of fit parameters, which is given
by the couplings and masses of the vector meson, $N_V = 4 + 3(N_s +N_v) +4(N_S+N_V)$,
with $N_{s/v} (N_{S/V})$ the number of the effective narrow (broad) isoscalar/isovector poles
and the $4$ represents the $\omega$ and $\phi$ couplings, minus the
number of constraints, given by $N_C = 4 + 6 + 1$, referring to the low-$t$,
the high-$t$ constraints and the neutron charge radius squared, respectively.
If the latter in not included, $N_C =10$. Putting pieces together, we
have in total $N_F = N_V - N_C = 3(N_s +N_v) +4(N_S+N_V) - 7$ or $N_F = 3(N_s +N_v) +4(N_S+N_V) - 6$ fit
parameters (including or excluding the $(r_E^n)^2$-constraint).

\section{Data Analysis and Uncertainties}\label{sec:error}
In this section, we briefly describe how the fits of the spectral functions
to data are performed and how the statistical and systematic errors can be
determined. 

First, the quality of the fits is measured by means of two different $\chi^2$ functions,
$\chi^2_1$ and $\chi^2_2$, which are defined as
\begin{align}
	\chi^2_1 &= \sum_i\sum_k\frac{(n_k C_i - C(t_i,\theta_i,\vec{p}\,))^2}{(\sigma_i+\nu_i)^2}~,
	\label{eq:chi1}\\
	\chi^2_2 &= \sum_{i,j}\sum_k(n_k C_i - C(t_i,\theta_i,\vec{p}\,))[V^{-1}]_{ij} (n_k C_j - C(t_j,\theta_j,\vec{p}\,))~,
	\label{eq:chi2}
\end{align}
where the $C_i$ are the experimental data at the points $t_i,\theta_i$ and the
$C(t_i,\theta_i,\vec{p}\,)$ are the theoretical values for a given FF parametrization
for the parameter values contained in $\vec{p}$.
For total cross sections and form factor data the dependence on $\theta_i$
is dropped. Moreover, the $n_k$ are normalization
coefficients for the various data sets (labeled by the integer $k$ and only used in the fits to
the differential cross section data in the spacelike region), while $\sigma_i$ and $\nu_i$ are 
their statistical and systematical errors, respectively. The covariance matrix
$V_{ij} = \sigma_i\sigma_j\delta_{ij} + \nu_i\nu_j$.
$\chi^2_2$ is used for those experimental data where statistical and systematical errors are given separately, otherwise 
$\chi^2_1$ is adopted. Furthermore, the $\chi^2$ of each data set is normalized by the number of data points in order to weight the various data sets without bias.

One also considers the {\em reduced} $\chi^2$, which is given by:
\beq
\chi^2_{\rm red} = \frac{\chi^2_i}{N_{D} - N_{F}}~, ~~i =1,2~,
\eeq
with $N_D$ the number of fitted data points and $N_F$ the
number of independent fit parameters, see Sec.~\ref{sec:con}.

As noted in Sec.~\ref{sec:con} the various constraints on the form factors can be implemented
algebraically (hard constraints) or by modifying the $\chi^2$ (soft constraints).
The latter type of constraints are implemented as additive terms to the
total $\chi^2$ of the following form
\beq
\chi^2_{\rm add.} = p\, [x-\langle x\rangle]^2 \,\exp\left(p\, [x-\langle x\rangle]^2\right)~,
\eeq
where $\langle x\rangle$ is the desired value and $p$ is a strength parameter,
which regulates the steepness of the exponential well and helps to stabilize the
fits~\cite{Belushkin:2006qa,Hammer:2006mw}. The fits are performed with {\em MINUIT} \cite{James:1975dr} in Fortran.

We now turn to the estimation of uncertainty. One method to estimate the fit (statistical) errors is 
the bootstrap procedure, see e.g. Ref.~\cite{Efron:1986hys}. 
One simulates a large number of data sets compared to the number of
data points by randomly varying the points in the original set within the given
errors assuming their normal distribution. Let us consider the radius extraction. In that
case, one fits to each of these data sets separately, extracts the radius from each fit
and consider the distribution of these radius values, which is sometimes denoted as
bootstrap distribution. The artificial data sets represent many real samples.
Therefore, this radius distribution emulates the probability distribution that one
would get from fits to data from a high number of measurements. The precondition for
using this method are independent and identically distributed data points. This is
fulfilled when  the $\chi^2$ sum does not depend on the sequential order of the
contributing points. For $n$ simulated data sets, the errors thus scale with $1/\sqrt{n}$.
However, to get a more realistic uncertainty, we exclude one percent of the data
points from the sample and so can determine the lowest and highest value of the
extracted radius. The same procedure can, of course, also be applied to the full
form factors. 

Another statistical tool to estimate the error intervals of our model parameters
is the Bayesian approach, see e.g. Ref.~\cite{Schindler:2008fh} and references therein. In contrast to the interpretation of probabilities in the
classical (also called frequentist) approach,
where the  probability is the frequency of an event to occur over a large
number of repeated trials, the Bayesian method
uses probabilities to express the current state of knowledge about the unknown parameters.
This approach enables dynamic updates to parameter estimates, accounting for both the parameter value and its associated uncertainty in a unified probabilistic framework.
The key ingredients to a Bayesian analysis are
the prior distribution, which quantifies what is known about the model parameters prior to
data being observed, and the likelihood function, which describes information about the
parameters contained in the data. The prior distribution and likelihood can be combined
to derive the posterior distribution by means of Bayes' theorem:
\begin{equation}
P(\mathrm{paras}|\mathrm{data})=\frac{P(\mathrm{paras})
	P(\mathrm{data}|\mathrm{paras})}{P(\mathrm{data})}~,
\end{equation}
where ``paras'' denotes the parameters and $P(a|b)$ is the conditional probability
that $a$ happens given $b$.

It is the main goal of a  Bayesian statistical analysis to obtain the posterior
distribution of the model parameters. The posterior distribution contains the total
knowledge about the model parameters after the data have been observed. From a
Bayesian perspective, any statistical inference of interest can be obtained through
an appropriate analysis of the posterior distribution. For example, point estimates of
parameters are commonly computed as the mean of the posterior distribution and interval
estimates can be calculated by producing the end points of an interval that correspond
with specified percentiles of the posterior distribution. A powerful and easy-to-implement
method to access the posterior distribution is the Markov Chain Monte Carlo (MCMC) algorithm. 
A systematic illustration of Bayesian analysis applications
can be found in Ref.~\cite{Wesolowski:2015fqa}. 
The analysis in \cite{Lin:2021umz} demonstrated that both bootstrap sampling and Bayesian simulation yield equivalent statistical errors for the form factors and radii. In the following, we will use the bootstrap procedure, as it is easier to implement for large data sets 
with a greater number of fit parameters. 

Next, we discuss the extraction of the {\em systematic} uncertainties, which is
always the most difficult task. Our strategy is similar to what was already done
in Ref.~\cite{Hohler:1976ax}, namely to vary the number of isoscalar and isovector
poles around the values corresponding to the best solution, where the total $\chi^2$
does not change by more than 1\%. An example of this is given in Tab.~\ref{tab:syst}
taken from Ref.~\cite{Lin:2021umk} where only the PRad data~\cite{Xiong:2019umf}
are considered. The best fit corresponds to 2 isoscalar and 2
isovector poles and the systematic errors in this case can be read off as
\cite{Lin:2021umk}
\beq
\delta (r_E^p)_{\rm syst.} = \pm 0.001~{\rm fm}~, ~~ \delta (r_M^p) = {}^{+0.018}_{-0.012}~{\rm fm}~.
\eeq
While the absolute $\chi^2$ does not change, the reduced one
worsens as the number of fit parameter increases. As expected, the systematic
error is larger for the magnetic radius as the electric FF dominates at low $Q^2$.
\begin{table}[t]
\centering  
\begin{tabular}{|ccc|cc|}
\hline
eff. poles & tot. $\chi^2$ & red. $\chi^2$  &   $r_E^p$~[fm]    &  $r_M^n$~[fm] \\
\hline
$2s+2v$*     &  88.5046   & 1.321   &   0.829    &   0.843 \\
$3s+2v$      &  88.5159   & 1.383   &   0.829    &   0.861 \\
$3s+3v$      &  88.5051   & 1.451   &   0.828    &   0.848 \\
$4s+3v$      &  88.5037   & 1.526   &   0.829    &   0.843 \\
$4s+4v$      &  88.5046   & 1.609   &   0.829    &   0.845 \\
$5s+4v$      &  88.5027   & 1.702   &   0.829    &   0.837 \\
$5s+5v$      &  88.5043   & 1.806   &   0.828    &   0.861 \\
\hline
\end{tabular}
\caption{Fit to the PRad data with varying numbers of isoscalar ($s$)
  and isovector ($v$) effective poles. Given are the total and the
  reduced $\chi^2$ and the resulting values for the proton radii. The *
  marks the best solution which defines the central values for the radii.
}
\label{tab:syst}
\vspace{-3mm}
\end{table}

\section{Nucleon Form Factor Results}

\begin{table}[htbp]
	\centering
	\renewcommand\arraystretch{1}
	\begin{tabular}{p{2.1cm}<{\centering}|p{2.1cm}<{\centering}|p{1.8cm}<{\centering}|p{2.4cm}<{\centering}|p{1.8cm}<{\centering}|p{2.0cm}<{\centering}}
		\hline
		\multicolumn{6}{c}{Experimental data}\\
		\cline{1-6}
		Region & Observables & Souce & $\abs{t}\ {\rm GeV^2}$ & number & References \\
		\hline
		\multirow{6}*{spacelike $t<0$}& \multirow{2}*{$d\sigma/d\Omega$} &MAMI &0.00384-0.977&1422 &~\cite{A1:2013fsc}\\
		&&PRad&0.000215-0.058&71&\cite{Xiong:2019umf}\\
		&$\mu_pG^p_E/G^p_M$&JLab&1.18-8.49&16&~\cite{Punjabi:2005wq,Puckett:2010ac,GEp2gamma:2010gvp,Puckett:2011xg}\\
		&$\mu_nG^n_E/G^n_M$&world&1.58-3.41&4&\cite{Belushkin:2006qa}\\
		&$G^n_E$&world&0.14-3.41&29&\cite{Riordan:2010id,Belushkin:2006qa} \\
		&$G^n_M$&world&0.071-10.0&49&\cite{Belushkin:2006qa}\\
		\cline{1-6}
		\multirow{4}*{timelike $t>0$}&$|{G^p_{\rm eff}}|$&world &3.52-20.25&153&\cite{BESIII:2021rqk,BESIII:2019hdp,BESIII:2019tgo,BESIII:2015axk,BaBar:2013ves,E835:1999mlt,Andreotti:2003bt}\\
		&$|G^n_{\rm eff}|$&world&3.52-9.49&32&\cite{BESIII:2021tbq,Druzhinin:2019gpo}\\
		&$|G^p_E/G^p_M|$&BaBar&3.52-9.0&6&\cite{BESIII:2021rqk}\\
		&$d\sigma/d\Omega$&BES\Rom{3}&3.52-3.8&10&\cite{BaBar:2013ves}\\
		\hline
	\end{tabular}
	\caption{Data base used in the dispersion-theoretical fits.}
	\label{tab:dbase}
\end{table}
In this section, we present a variety of physics results based on a recent state-of-the-art dispersion theoretical analysis of the world data set~\cite{Lin:2021xrc}.
Specifically, we discuss fits that include the differential cross section from the electron-proton elastic 
scattering, the proton FF ratio from the polarization transfer experiments and neutron FF data in the space-like region, 
as well as the effective FF for both the proton and neutron, and the proton FF ratio in the time-like region.
The data base fitted in Ref.~\cite{Lin:2021xrc} is presented in Tab.~\ref{tab:dbase}.
The best fit is found to consist of 3 narrow poles in the isoscalar channel $(s)$
and 5 narrow poles in the isovector channel $(v)$ below the nucleon-nucleon threshold 
and $3s+3v$ broad poles above the threshold, that is, $N_s=3$, $N_v=5$ and $N_S=N_V=3$.
$\chi^2/{\rm d.o.f}=1.223$ is obtained for this best fit with the $(r_E^n)^2$-constraint included.
Note that there are 33 additional normalization constants for the MAMI and PRad data in the spacelike region. These are discussed in detail in
Ref.~\cite{Lin:2021umz}. We remark that fits with fixed normalizations
lead to the same results but larger $\chi^2/$dof. 
The vector meson parameters for the best fit can be found in the supplemental materials of Ref.~\cite{Lin:2021xrc}.

\subsection{Spacelike Form Factors}
\label{sec:newfits}

\begin{figure}[htbp]
    \centering
    \begin{minipage}[c]{0.45\textwidth} 
        \centering
        \includegraphics[height=0.6\textheight]{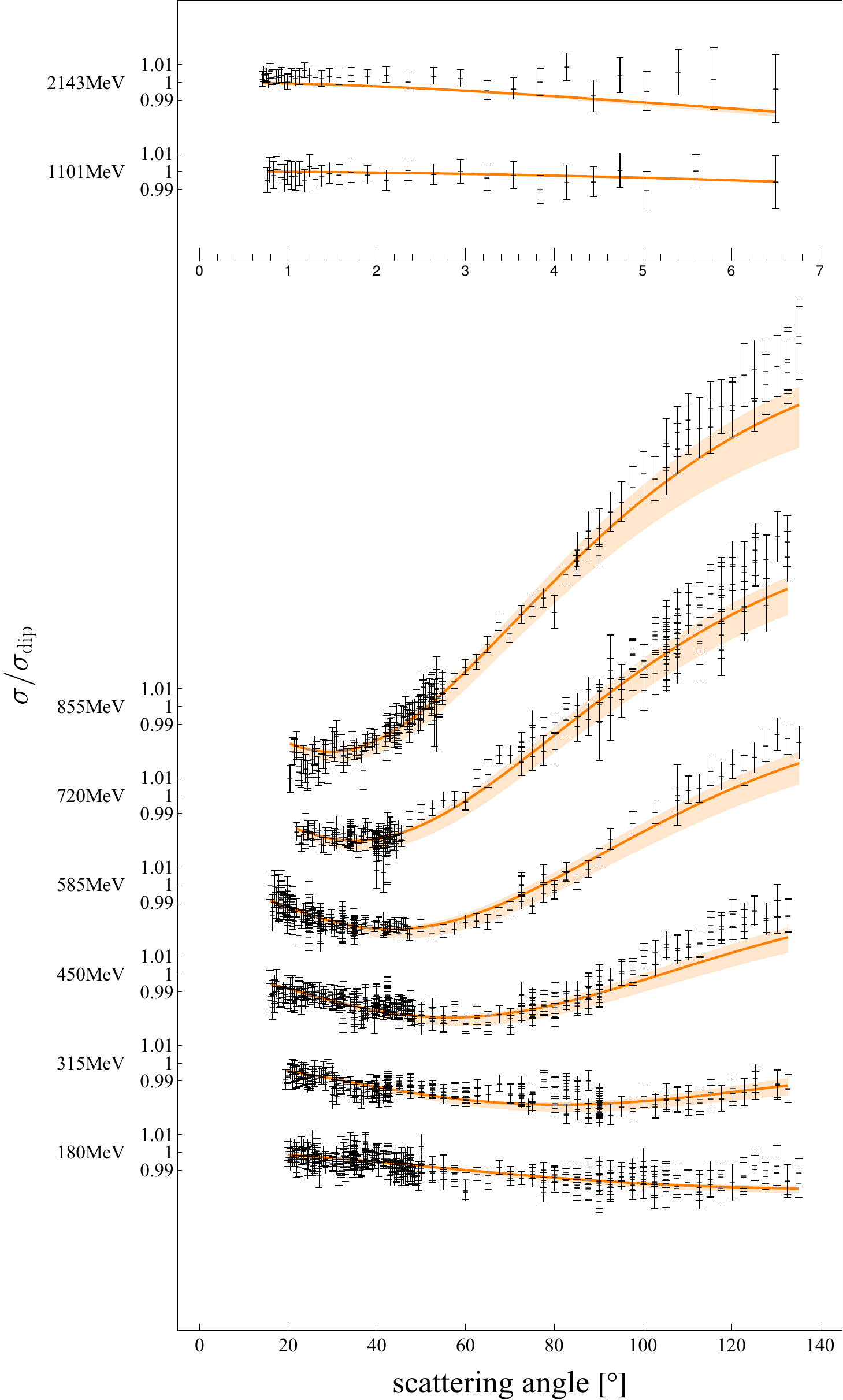} 
        \label{fig:long}
    \end{minipage}%
    \hfill
    \begin{minipage}[c]{0.45\textwidth} 
        \centering
         \begin{tabular}{@{}c@{}}
            \includegraphics[width=0.98\linewidth]{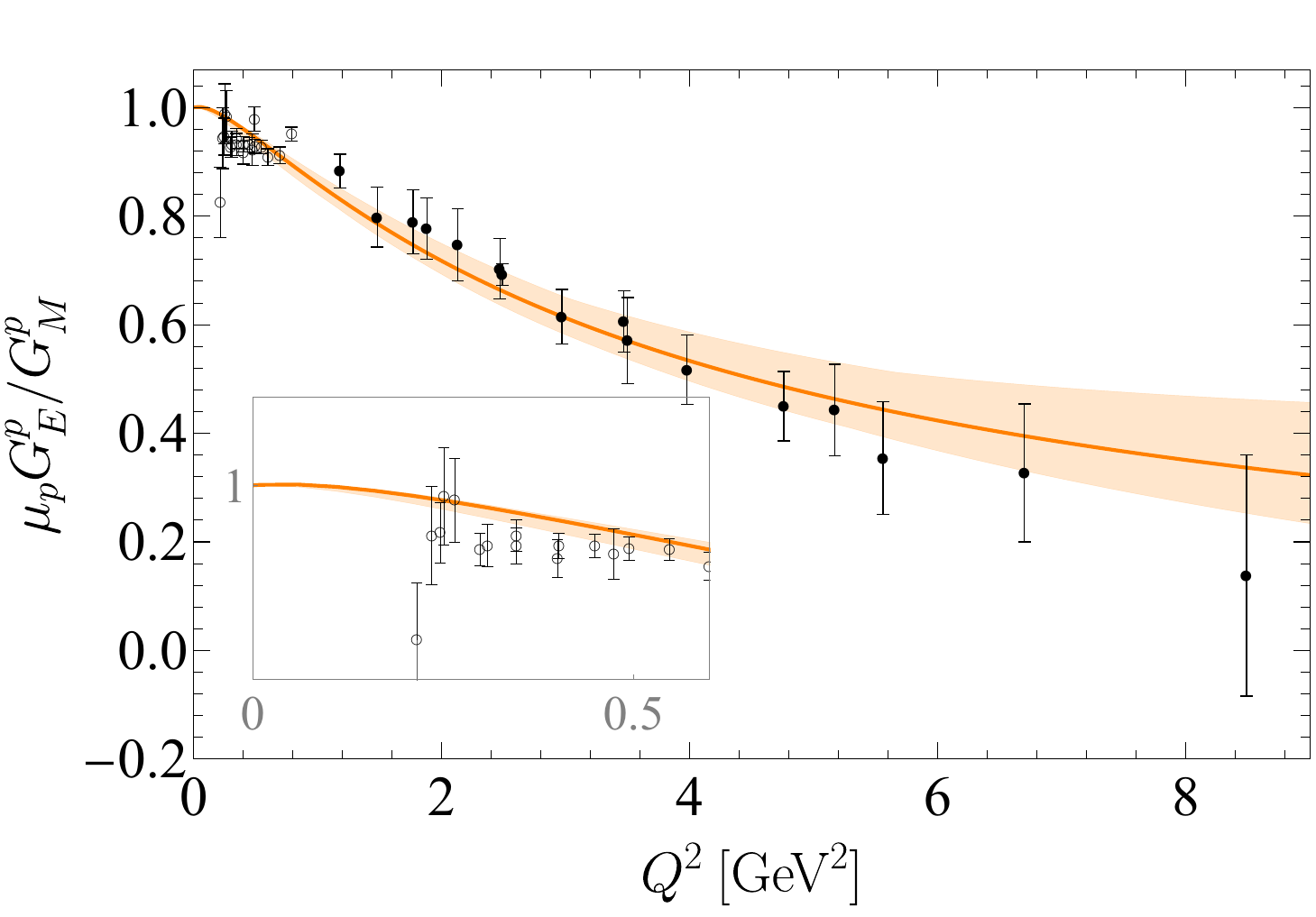} \\
            \includegraphics[width=\linewidth]{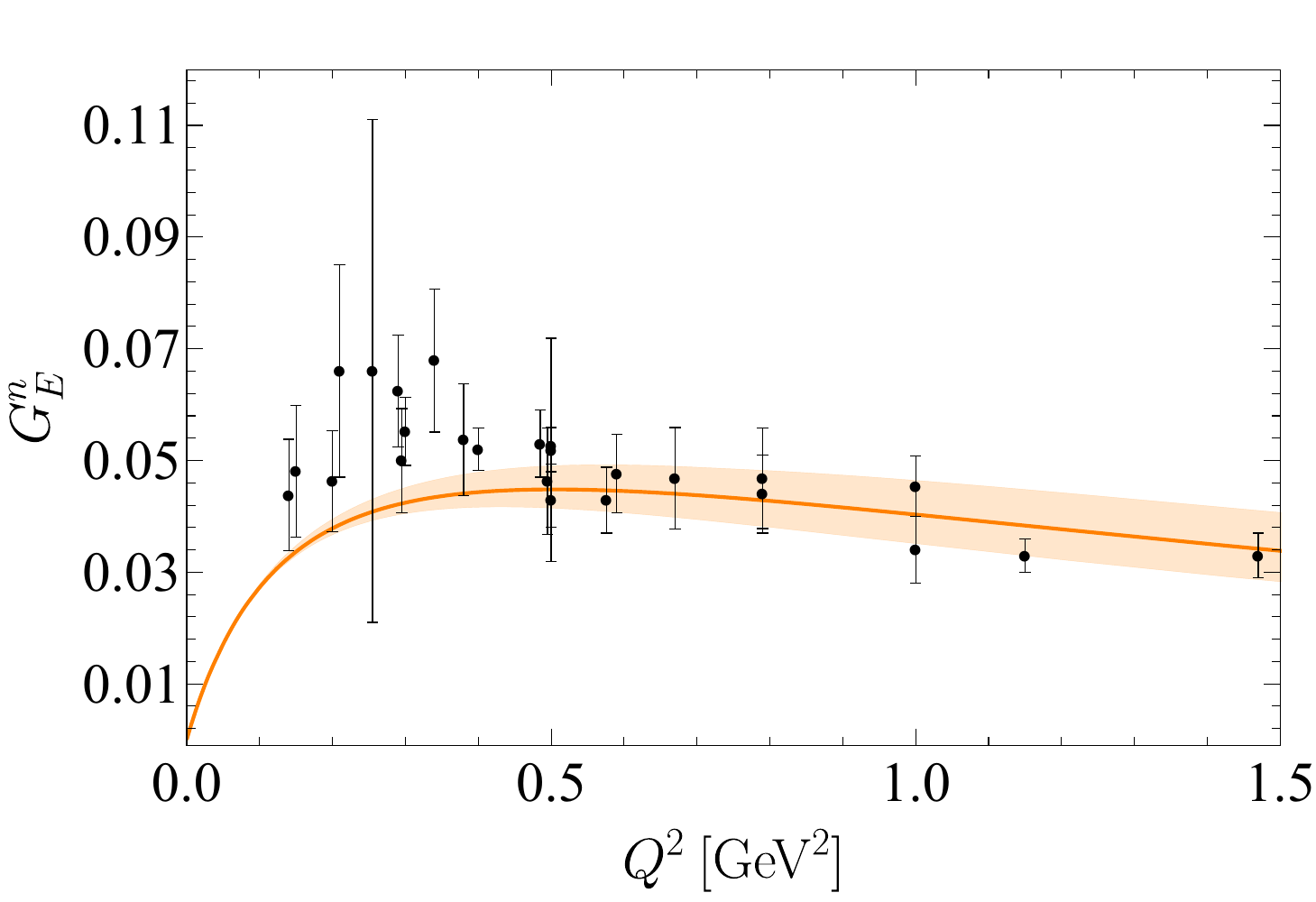} \\
            \includegraphics[width=0.98\linewidth]{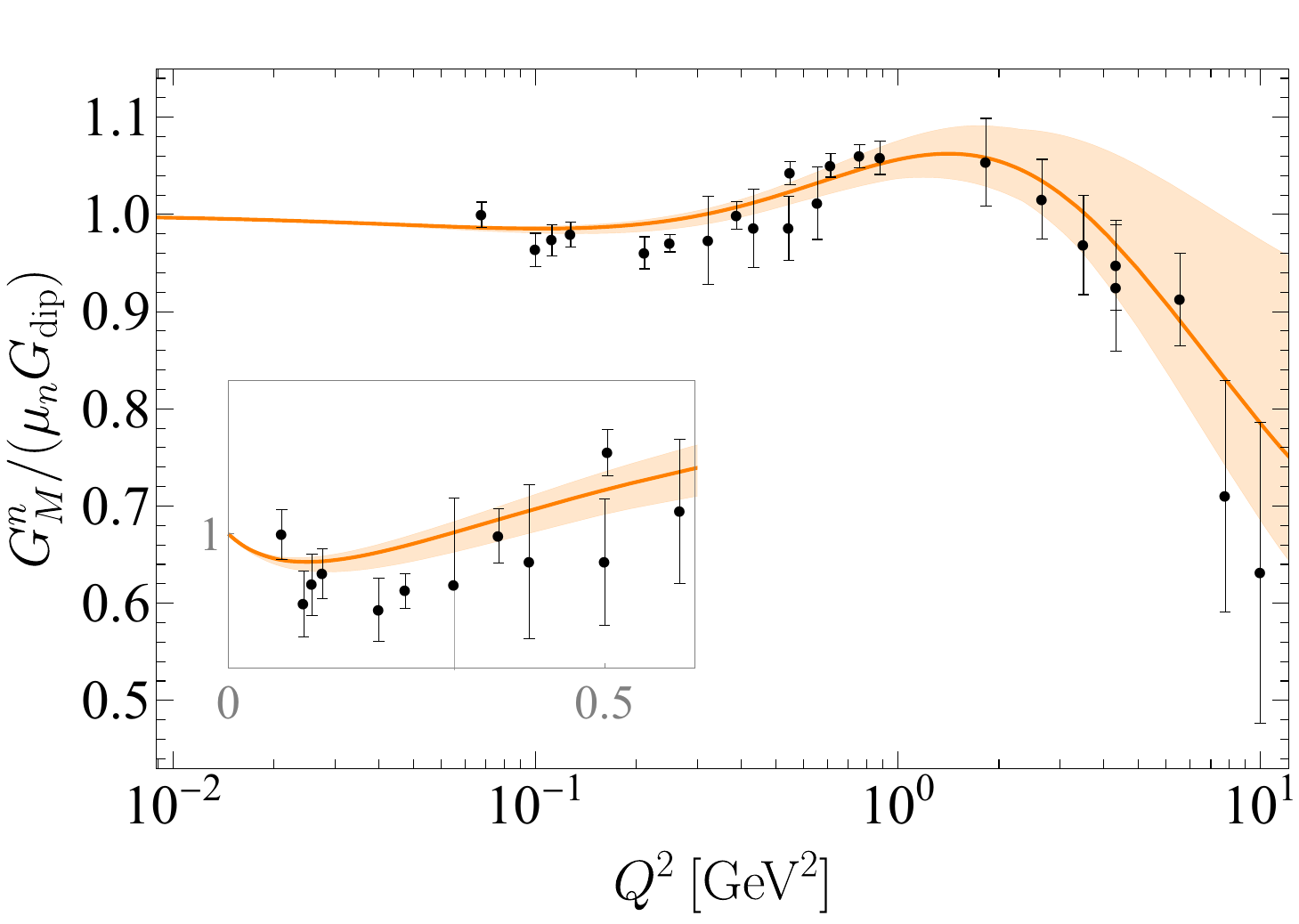}
        \end{tabular}
    \end{minipage}
\caption{Complete fit to space- and timelike data
	with bootstrap error (shaded band) compared to the $ep$ cross section data from PRad (left upper panel)
	and MAMI (left lower panel), the JLab data for $\mu_p G_E^p/G_M^p$(right top panel), 
	the neutron electric form factor data (right middle panel)
	and the neutron magnetic form factor data (right bottom panel) at spacelike momentum transfer.
	Fitted data are depicted by closed symbols. The data for $|t|<1$~GeV$^2$ (open symbols, see also the inset) are
	shown for comparison only. The colored bands give the uncertainty due to the bootstrap procedure. Systematical uncertainties
	are not shown.
	\label{Fig: spacelike}
}
\end{figure}
In Fig.~\ref{Fig: spacelike}, we show our best fit compared to
the experimental data for the $ep$ cross section data from PRad (left upper panel)
and MAMI (left lower panel), $\mu_p G_E^p/G_M^p$ from JLab (right top panel), 
the neutron electric form factor (right middle panel)
and the neutron magnetic form factor (right bottom panel) at spacelike momentum transfer.
Note that for the proton case, we fit to the $ep$ cross section data with $Q^2<1\,$GeV$^2$,
incorporating with the proton form factor ratio data with $Q^2>1\,$GeV$^2$.
As in earlier fits~\cite{Lorenz:2014yda,Lin:2021umk}, the data for the
proton form factor ratio $\mu_PG_E^p/G_M^p$ for $Q^2<1\,$GeV$^2$, which
do not participate in the fit, are well described, see the inset in the right top panel in
Fig.~\ref{Fig: spacelike}. This points towards consistency between the
two-photon corrected cross section data and the ratio data, that are
not affected by such corrections. 

Moreover, a decreasing behavior of $G_M^n/(\mu_n G_{\rm dip})$ and $\mu_p G_E^p/G_M^p$ at large $|t|$ in the spacelike region is explicitly enforced
in order to get a good description over the full range of momentum
transfers. It turns out that a zero crossing of $\mu_p G_E^p/G_M^p$
is disfavored by the combined analysis of space- and timelike
data, while some measurements suggest
a zero crossing of this ratio around $t \approx -10$~GeV$^2$
\cite{Arrington:2011kb}. Thus, data at higher momentum transfer than shown in the figure
are required to settle this issue. We further remark that as in the
earlier fits to the spacelike data only, the onset of perturbative
QCD barely sets in at the highest momentum transfers probed.

Another facet of the spacelike FFs is the long-range part of the Breit-frame
charge and magnetization distributions that follows from the
Sachs form factors and
can be interpreted in terms of a ``pion cloud'' and some additional
short-range contributions from the $\rho$ and other short-ranged
physics. However, we emphasize that this separation is scale-dependent
and thus not unique~\cite{Hammer:2003qv,Meissner:2007tp}. 
We will come back to this issue in Sec.~\ref{sec:picloud}.

\subsection{Nucleon Radii}
Now, let us move to the nucleon radii. The radii extracted from the combined fits in \cite{Lin:2021xrc} are
\begin{equation}
	r_E^p = 0.840^{+0.003}_{-0.002}{}^{+0.002}_{-0.002}\,{\rm fm}, \quad
	r_M^p = 0.849^{+0.003}_{-0.003}{}^{+0.001}_{-0.004}~{\rm fm}, \quad
	r_M^n = 0.864^{+0.004}_{-0.004}{}^{+0.006}_{-0.001}~{\rm fm},
	\label{eq:radii_fin}
\end{equation}
where the first error is statistical (based on the bootstrap procedure) 
and the second one is systematic (based on the variations in the spectral functions).
These values are in good agreement with previous high-precision analyses
of spacelike data alone \cite{Lin:2021umk,Lin:2021umz} and have comparable errors. 

Alternative information on the
proton charge radius can be obtained from Lamb shift measurements in electronic
as well as muonic hydrogen, see e.g. the reviews~\cite{Pohl:2013yb,Karr:2020wgh,Gao:2021sml}.
The proton radius puzzle—marked by a striking discrepancy between 
the proton charge radius extracted from muonic hydrogen spectroscopy~\cite{Pohl:2010zza,Antognini:2013txn} 
and the value averaged from electron scattering and ordinary hydrogen
spectroscopy~\cite{Mohr:2008fa} has driven extensive experimental efforts 
in elastic electron-proton scattering over the past decade. 

\begin{figure}[htbp]
	\centering
	\includegraphics[width=0.9\textwidth]{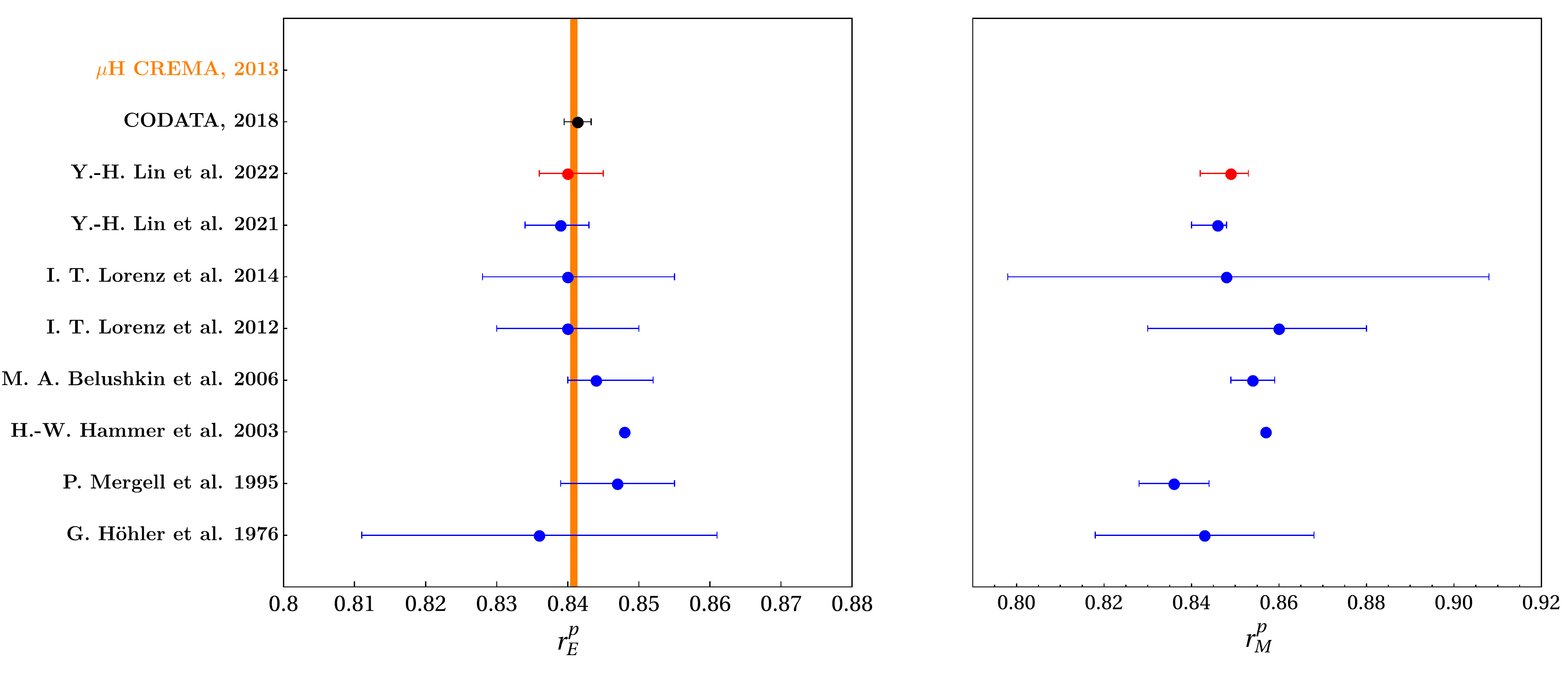}
	\caption{Comparison of the proton electric (left) and magnetic (right) radii determined by various dispersion-theoretical extractions. The y-axis represents the date and first author of the corresponding work, see Ref.~\cite{Lin:2021umz} for the relevant references. The orange band shows the latest radius extraction from the muonic hydrogen~\cite{Antognini:2013txn}.
		\label{fig:rpe}}
	\vspace{-3mm}
\end{figure}
For the proton charge radius $r_E^p$, we compare various dispersion-theoretical extractions from a historical perspective in Fig.~\ref{fig:rpe}. Note that here we only consider those dispersion-theoretical analyses that include the two-pion continuum explicitly in their spectral functions that is found to play a crucial role in the nucleon isovector form factors, see Refs.~\cite{Lin:2021umz} for the details. It is remarkable that the dispersion-theoretical analyses always provided a consistent and robust proton charge radius in agreement with the spectroscopic values from muonic hydrogen~\cite{Pohl:2010zza,Antognini:2013txn}. 
\begin{figure}[htbp]
	\centering
	\includegraphics[width=0.9\textwidth]{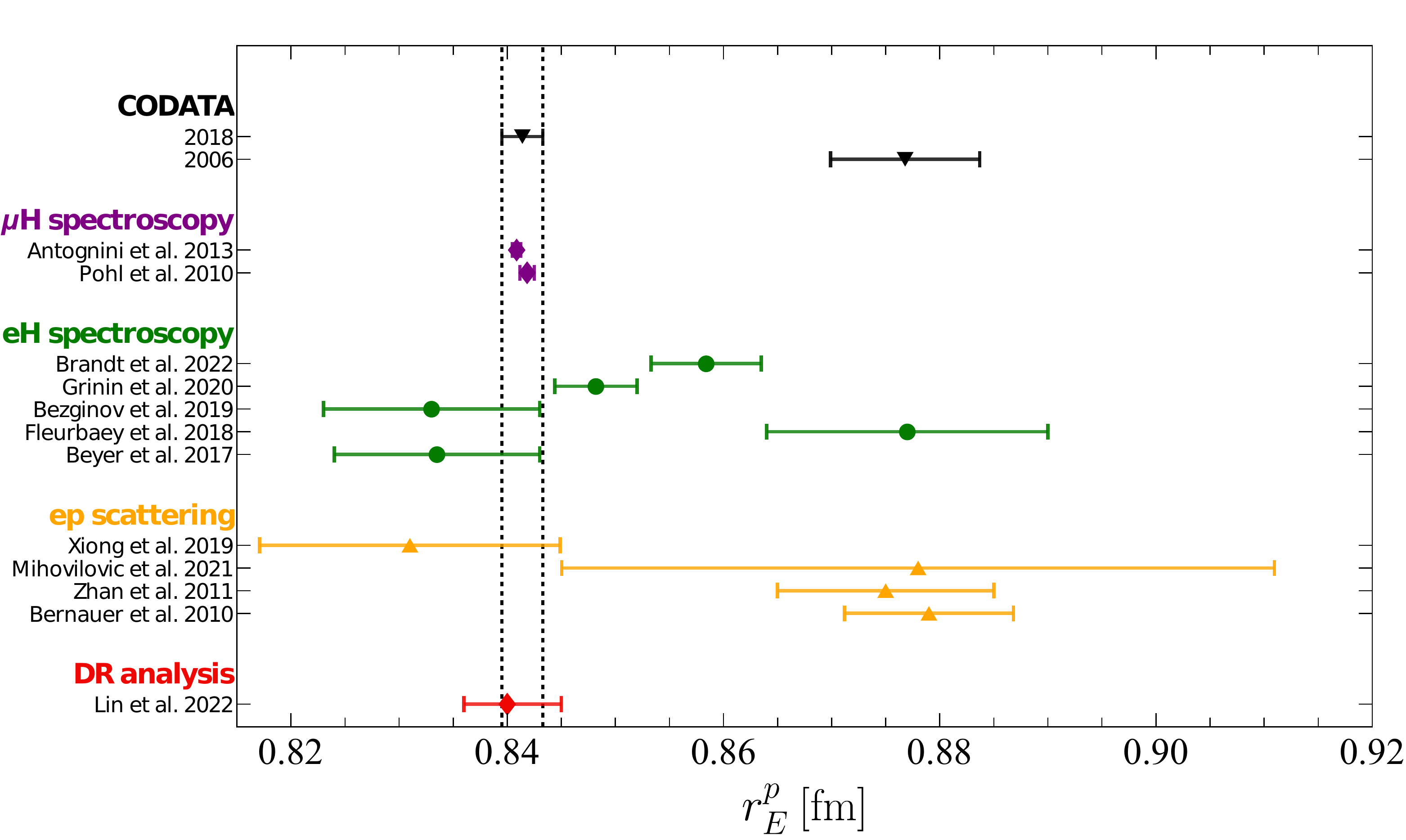}
	\caption{Comparison of the proton charge radius extracted in Ref.~\cite{Lin:2021xrc} and other recent determinations. The y-axis indicates the process in which the proton charge radius was extracted as well as the date and first author of the corresponding work, see Ref.~\cite{Lin:2021umz} for the relevant references.
		\label{fig:rpeoverview}}
	\vspace{-3mm}
\end{figure}
In Fig.~\ref{fig:rpeoverview}, we list the most recent experimental determinations of the proton electric radius. Agreement on the proton charge radius has been achieved by the measurements from $ep$ scattering, $ep$ spectroscopy and the $\mu p$ spectroscopy. As a consequence, the value collected in CODATA was updated in 2018 and the 2022 value is $0.84075(64)$\,fm~\cite{CODATAnew,Mohr:2024kco} which agrees quite well with the dispersion theoretical determination of Ref.~\cite{Lin:2021xrc}. 
Moreover, the Zemach radius~\cite{Zemach:1956zz}
\beq
r_Z = -\frac{4}{\pi}\int_0^\infty \frac{dQ}{Q^2}\left[\frac{G_E(Q^2)G_M(Q^2)}{1+\kappa}-1\right]\,,
\label{eq:Zemach}
\eeq
and the third Zemach moment $\langle r^3 \rangle_{(2)}$, are obtained as
\begin{equation}
	r_z = 1.054^{+0.003}_{-0.002}{}^{+0.000}_{-0.001}\,{\rm fm},\quad	\langle r^3\rangle_{(2)} = 2.310^{+0.022}_{-0.018}{}^{+0.014}_{-0.015}~{\rm fm}^3.
	\label{eq:radii_zemach}
\end{equation}
These values are in good agreement with Lamb shift and hyperfine splittings in muonic hydrogen \cite{Antognini:2013txn}. 

While the electric radius of the proton has attracted
much attention in the last decade, this is not true for its magnetic counterpart.
The magnetic radius is not probed directly in the Lamb
shift in electronic or muonic hydrogen and thus all existing
information comes from electron scattering experiments. The $ep$ cross section, however,  is dominated by the electric form factor for small momentum
transfer. Thus the magnetic radius $r_M$ is more sensitive
to larger momentum transfers, and it is not known experimentally with the same precision as $r_E$.
The dispersion-theoretical values of the proton magnetic radius $r_M^p$ were consistently bigger than 0.83~fm and slightly larger than $r_E^p$, as seen in Fig.~\ref{fig:rpe}. 
In stark contrast, the analysis of the A1 collaboration~\cite{Bernauer:2010wm} yielded
a significantly smaller value of $r_M = 0.777(13)_{\rm stat.}(9)_{\rm syst.}(5)_{\rm model}(2)_{\rm group}$~fm, including in addition to
statistical and systematic uncertainties also some uncertainties from the
fit model and differences between the two model groups used in the analysis.
However, looking at the corresponding magnetic form factor $G_M(Q^2)$, it shows a pronounced bump-dip structure for
momentum transfers $0\leq Q^2 \leq 0.3$~GeV$^2$. Such a structure is at odds with
unitarity and analyticity \cite{Hammer:2006mw,Meissner:2007tp}. 
So is there other information available that could help
to clarify this issue? Indeed, lattice QCD calculations at physical pion masses are available.
The latest lattice value for $r_M$ given in Ref.~\cite{Djukanovic:2023jag}, $r_M = 0.8111(89)$~fm,
roughly corresponds to a 4$\sigma$ deviation from the
dispersive value. Note, however, that the electric radius
in that work also comes out rather small, $r_E = 0.820(14)$~fm.
The recent results by the PACS collaboration which include finite lattice spacing effects \cite{Tsuji:2023llh} also feature a small
magnetic radius, but they have larger errors and are consistent with the dispersive value.
These conflicting determinations of the proton magnetic radius appear to reveal a "new proton radius puzzle"~\cite{Lin:2023fhr}.
Assuming that the proton electric radius is known now, a very precise determination of the Zemach radius, Eq.~(\ref{eq:Zemach}),
which also enters into the Lamb shift,
would give another independent determination of $r_M$ that could help to clarify this issue.

\subsection{Timelike Form Factors}
\begin{figure}[htbp]
	\centering
	\includegraphics[width=0.45\textwidth]{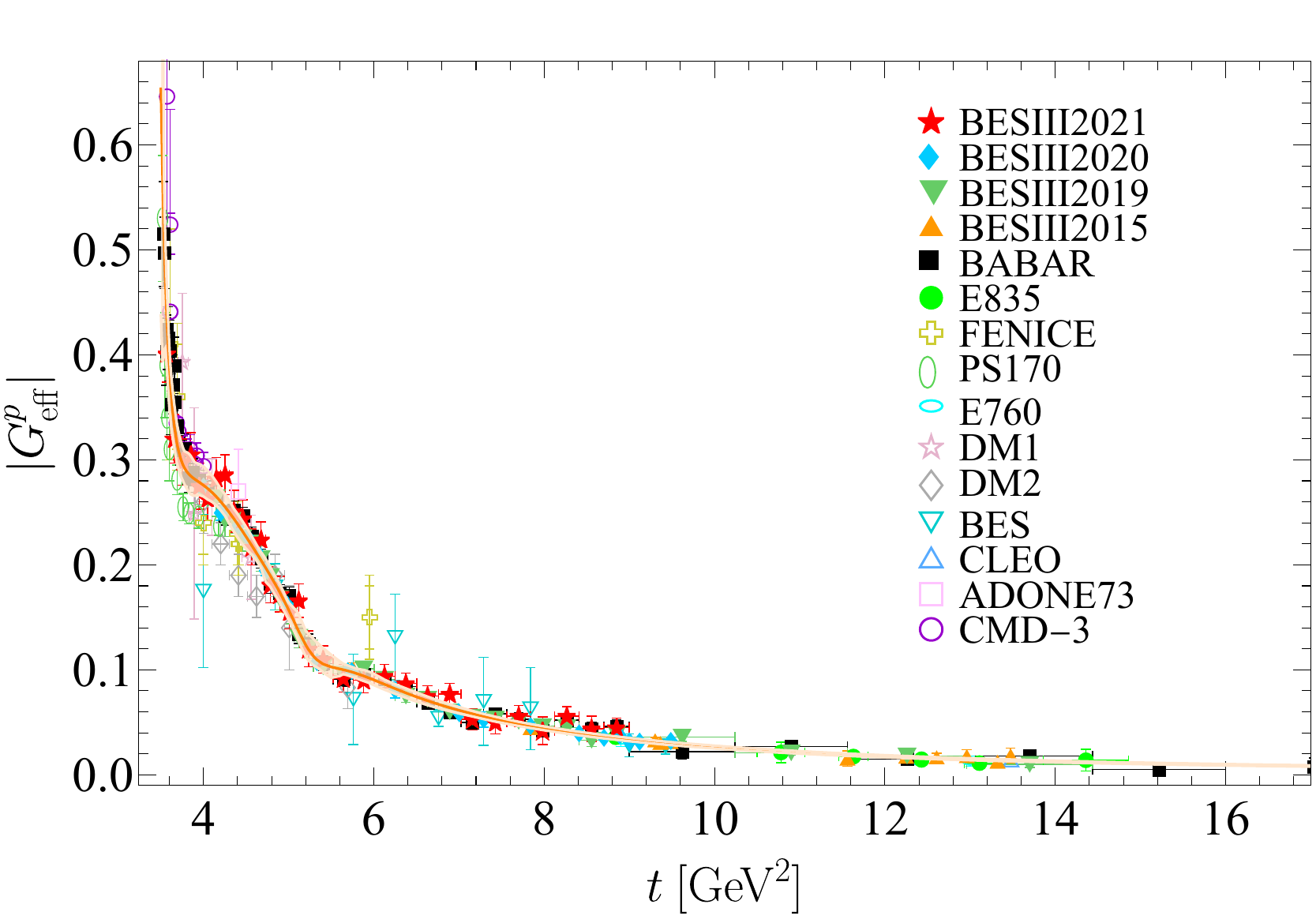}\
	\includegraphics[width=0.47\textwidth]{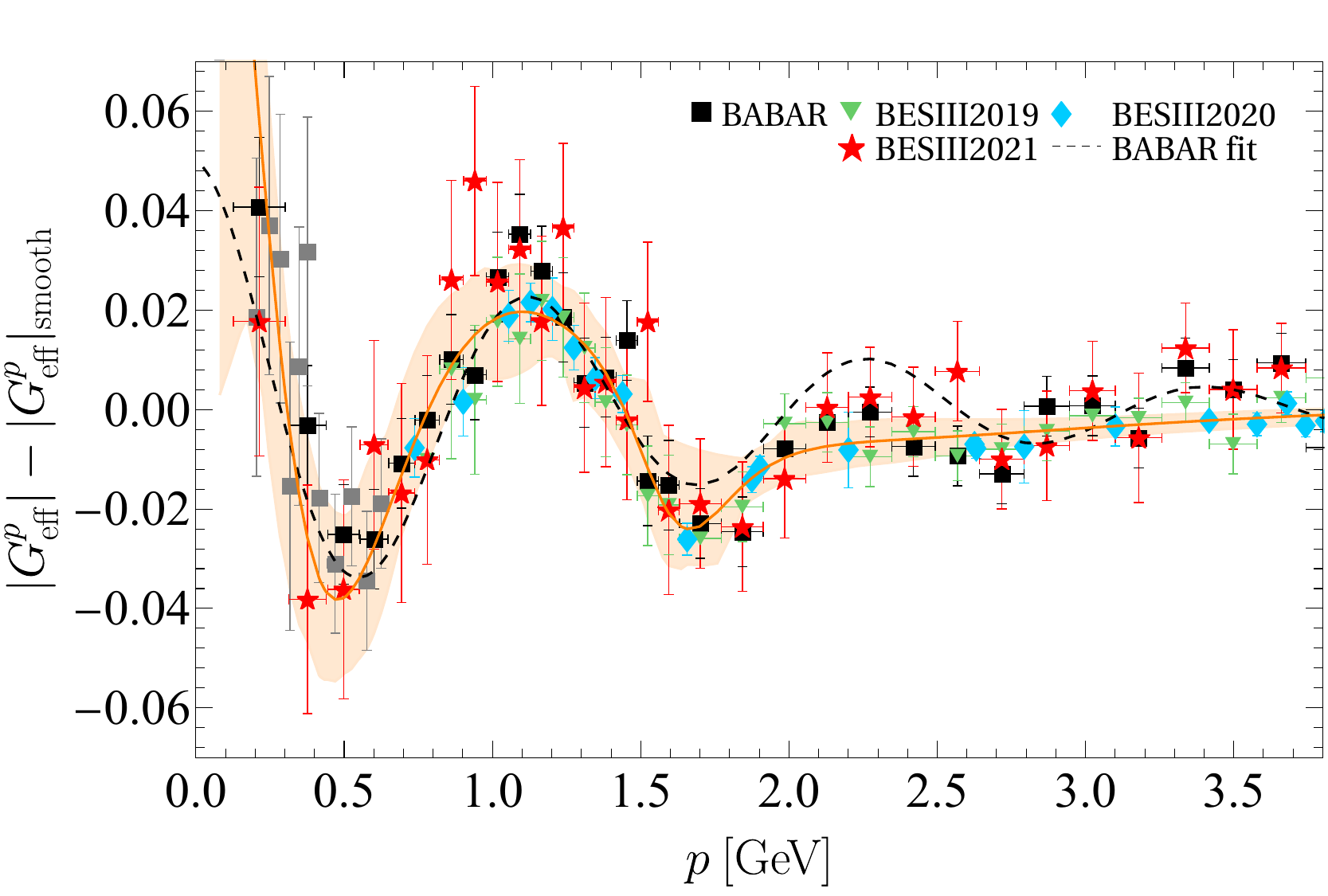}
	\caption{Complete fit to space- and timelike data
		with bootstrap error (shaded band)
		compared to data for $|G_{\rm eff}|$
		of the proton (left panel) and the oscillatory behavior in detail (right panel).
		Fitted data are depicted by closed symbols; data given
		by open symbols are shown for comparison
		only (see Ref.~\cite{Lin:2021xrc} for explicit references). $\left| G_{\rm eff}^p\right|_{\rm smooth}=7.7/(1+t/14.8)/(1-t/0.71)^2$~\cite{BaBar:2013ves}. The black dashed line in the right plot show the phenomenological fit to BABAR data with the formula $F_p\equiv \left| G_{\rm eff}^p\right|-\left| G_{\rm eff}^p\right|_{\rm smooth}=A \exp(-B p)\cos(C p+D)$ proposed in Ref.~\cite{Bianconi:2015owa}. Here, $p$ is the relative momentum of the proton.}
	\label{Fig: geffp}
\end{figure}

\begin{figure}[htbp]
	\centering
	\includegraphics[width=0.45\textwidth]{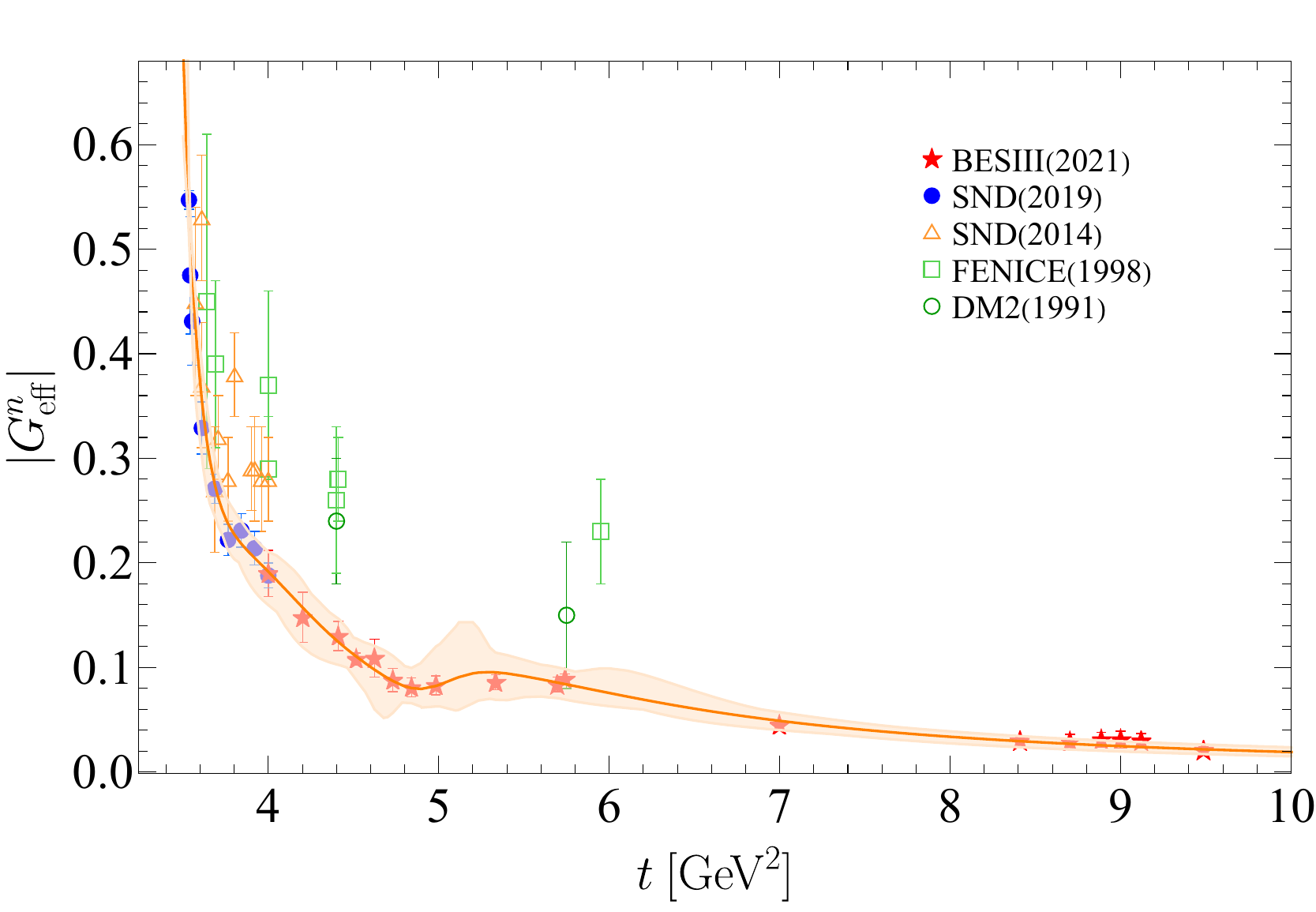}\
	\includegraphics[width=0.47\textwidth]{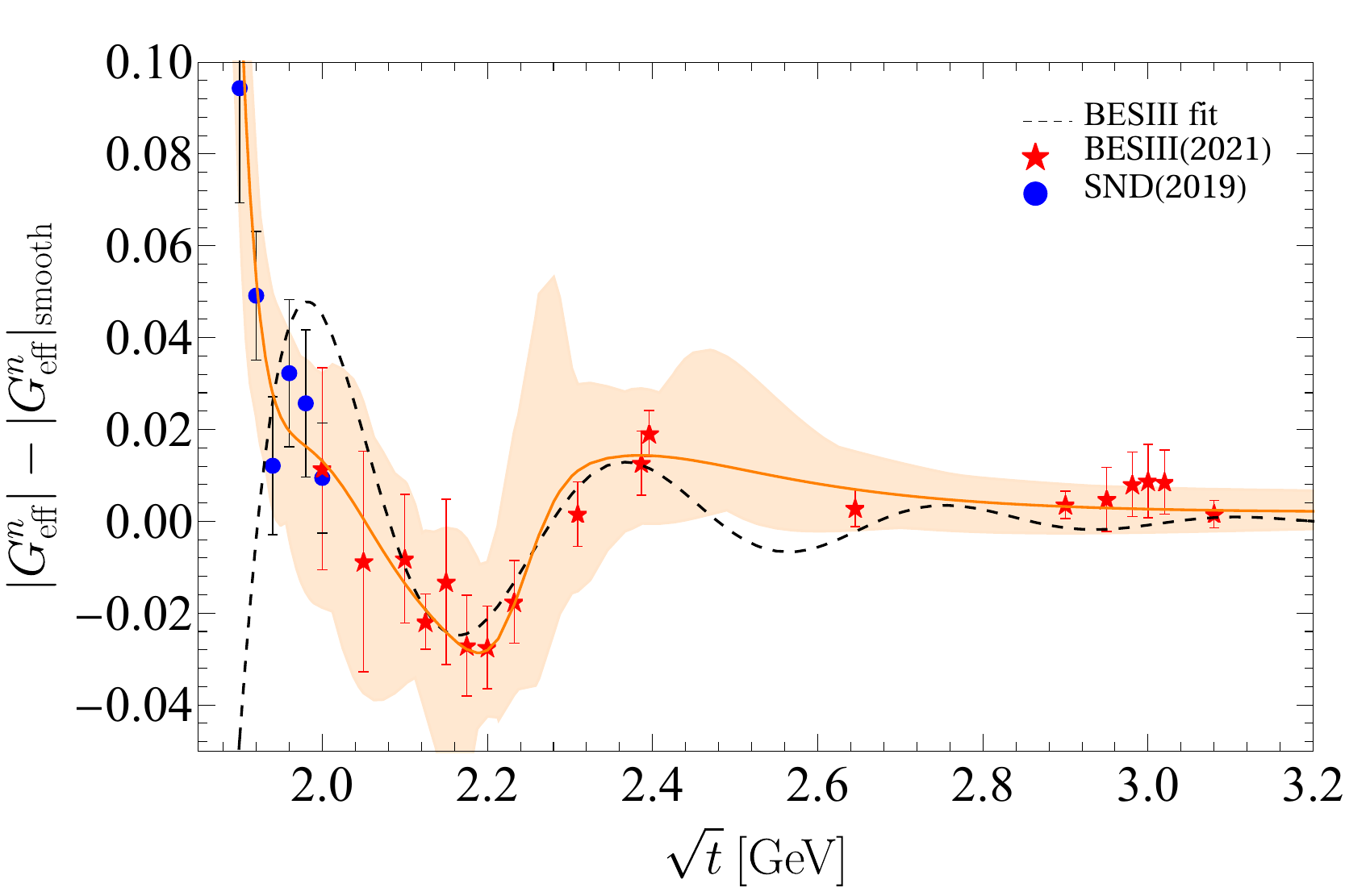}
	\caption{Complete fit to space- and timelike data
		with bootstrap error (shaded band)
		compared to data for $|G_{\rm eff}|$
		of the neutron (left panel) and the oscillatory behavior in detail (right panel).
		Fitted data are depicted by closed symbols; data given
		by open symbols are shown for comparison
		only (see Ref.~\cite{Lin:2021xrc} for explicit references). $\left| G_{\rm eff}^n\right|_{\rm smooth}=4.87/(1+t/14.8)/(1-t/0.71)^2$~\cite{BESIII:2021tbq}. The black dashed line in the right plot show the phenomenological fit to BESIII data with the formula~\cite{Bianconi:2015owa} $F_n\equiv \left| G_{\rm eff}^n\right|-\left| G_{\rm eff}^n\right|_{\rm smooth}=A \exp(-B p)\cos(C p+D)$, with $p$ is the relative momentum of the neutron.}
	\label{Fig: geffn}
\end{figure}

In Fig.~\ref{Fig: geffp} and Fig.~\ref{Fig: geffn}, we show our best fit compared to
the experimental data for $|G_{\rm eff}|$ of the proton and the neutron, respectively.
We obtain a good description of the timelike data for $|G_{\rm eff}|$. 
With $3s+5v$ 
below-threshold narrow poles and $3s+3v$ above-threshold broad
poles, we were able to reproduce both the visible near-threshold enhancement
of the proton and the neutron timelike form factor (after subtraction of
the electromagnetic final-state interaction in the proton case), first seen
by the PS170 collaboration at LEAR~\cite{Bardin:1994am}, 
and the prominent oscillations in $|G_{\rm eff}|$
between the threshold at $t = 4m^2$ and $t\approx 6$~GeV$^2$.
These poles also generate the imaginary part
of the form factors in the physical region. 
Alternatively, these structures can also be generated by including
contributions from triangle diagrams with $\Delta\bar{\Delta}$
and $(\Delta\bar{N}+{\rm h.c.})$ intermediate states, see, e.g.,
Ref.~\cite{Lorenz:2015pba}. In principle, these contributions
are fixed. However, the corresponding coupling constants are poorly
known and a perturbative treatment of these contributions
is questionable. For further discussion, see Ref.~\cite{Bianconi:2015owa}. 
Alternative interpretations of the oscillatory behavior of the timelike FFs of the nucleon are given in Refs.~\cite{Yang:2022qoy,Qian:2022whn,Yang:2024iuc,Rosini:2024lld}.

\subsection{Remarks on the Pion Cloud}\label{sec:picloud}
In the discussion of hadron properties the role of the pion cloud is frequently discussed.
Long before the discovery of QCD, it was realized that pion-nucleon scattering data require a large coupling constant. As a consequence of this strong coupling, many virtual mesons – Yukawa's pions – were expected to be associated with the nucleon, the pion cloud \cite{Schweber:1955zz}. Many of these ideas have survived until today, but now we know that low-energy QCD is governed by the spontaneous breakdown of its chiral symmetry (for the light quarks) with the pion
taking over a special role as a (Pseudo-)Goldstone boson. In this context, the pionic contribution
to the nucleon (or more generally baryon) structure, is called "pion cloud". While it is an important part of
nucleon structure, in general it is not possible to uniquely and unambiguously define the contribution of the pion cloud to a given observable. An evaluation of the two-pion contribution to the nucleon electromagnetic form factors by use of dispersion relations and chiral perturbation theory to underline this point was given in Ref.~\cite{Hammer:2003qv}.

Here we discuss the resolution dependence of the pion cloud  in the framework of chiral perturbation theory (ChPT) \cite{Meissner:2007tp}. To be precise, we consider a single nucleon. In baryon ChPT a nucleon typically emits a pion, this energetically forbidden 
$\pi N$ intermediate state lives for a short while before the pion is reabsorbed by the nucleon, in
accordance with the uncertainty principle. This mechanism generates the pion cloud of the nucleon, which in ChPT can be put on the firm ground of field theoretical principles. However, such loop contributions are in general scale-dependent, such that it is not possible
to unambiguously define the pion cloud. As a consequence,
the the concept of the pion cloud is resolution dependent.

To illustrate this fact, we consider the isovector Dirac radius of the proton  as an example \cite{Bernard:2003rp}. The first loop contributions appear at third order in the chiral expansion, leading to
\beq
\langle r^2\rangle_1^v =\left[0.68- (0.47\,\mathrm{GeV}^2)\bar{d}(\mu)+0.52 \ln (\mu/\mathrm{GeV})\right]\,\mathrm{fm}^2\,,
\eeq
where $\bar{d}(\mu)$ is a pion-nucleon low-energy constant that parameterizes the nucleon core contribution.
We emphasize that infinitely many combinations of $(\mu, \bar{d}(\mu))$ reproduce the empirical result $\langle r^2\rangle_1^v = 0.585$~fm$^2$ \cite{Belushkin:2006qa}, e.g.
(1~GeV, 0.20~GeV$^{-2}$), (0.835~GeV, 0.0 GeV$^{-2}$), and (0.6~GeV, $-0.37$~GeV$^{-2}$).
Even the sign of the core contribution to the radius can change within a reasonable range typically used for the scale  $\mu$. Physical intuition tells us that the value for the coupling $d$ should be negative such that the nucleon core gives a positive contribution to the isovector Dirac radius, but field theory shows that for (not unreasonable) regularization scales above $\mu = 835$~MeV this need not be the case. In essence, only the
sum of the core and the cloud contribution constitutes a meaningful quantity that should be discussed. This observation holds for any observable - not just for the isovector Dirac radius discussed here. Consequently, an unambiguous extraction of the pion cloud contribution is not possible.

\section{Hyperon form factors}

The successful dispersive treatment of the nucleon EMFFs also holds much promise for a model-independent description of the electromagnetic structure of hyperon states. These form factors are much less well-known compared to the nucleon case and experimental data are only available in the timelike region \cite{Schonning:2023hnv}, for a recent short review, see Ref.~\cite{Dai:2024lau}.
The authors of Ref.~\cite{Granados:2017cib} considered once-subtracted dispersion relations for the electromagnetic Sigma-to-Lambda transition form factors and expressed these in terms of the pion EMFF and the two-pion-Sigma-Lambda scattering amplitudes. They predicted the electromagnetic Sigma-to-Lambda transition form factors and investigated the role of pion rescattering and the role of the explicit inclusion of the decuplet baryons in three-flavor chiral perturbation theory (ChPT). In \cite{Lin:2022dyu}, the dispersion theoretical determination of the electromagnetic Sigma-to-Lambda transition form factors was extended to include the $K\bar{K}$ intermediate state in $\pi\pi-K\bar{K}$ coupled-channel treatment in SU(3) ChPT. This resulted in a shift of the electric Sigma-to-Lambda transition form factor $G_E$, while the magnetic form factor $G_M$ stays essentially unchanged. At present, the dispersion theoretical determination of electromagnetic Sigma-to-Lambda transition form factors suffers from sizeable uncertainties due to the poor knowledge of certain low-energy constants in SU(3) ChPT. The precise determination of this three-flavor LEC from the future experiments will be helpful to pin down the transition form factor. The electromagnetic form factors of the transition from the spin-3/2 $\Sigma$ to the $\Lambda$ hyperon were considered in Ref.~\cite{Junker:2019vvy}. Moreover, dispersion theory has been used to analyze the full set of cross section data for the reaction $e^+e^- \to \Lambda\bar{\Lambda}$ \cite{Lin:2022baj}. As more experimental data on the hyperon electromagnetic structure become available, dispersion theoretical methods provide a powerful tool to analyze these data and to predict space-like form factors from experimental data in the time-like region. They also
provide valuable insights in the underlying mechanisms.

\section{Conclusions}
\label{sec:conclusions}

This chapter reviews the status of baryon form factors with a special focus on the nucleon EM form factors which are known best.
In particular, we discuss the dispersion-theoretical
approach to the nucleon EM form factors as well as radii and highlight recent progress in the field. We emphasize that this approach has matured and become a precision tool to analyze electron scattering and form factor ratio data.
We stress that DRs have consistently found a
small proton charge radius, $r_E^p \simeq 0.84\,$fm, with a slightly larger proton magnetic radius, $r_M^p \simeq 0.85\,$fm. 
Regarding the latter, we point out that there are a number of conflicting determinations, which could be regarded as a new "proton radius puzzle".
We have also discuss our
present understanding of the physics in both the space- and time-like regions. 
Firstly, the combined analysis of space- and timelike
data disfavors a zero crossing for the proton FF ratio $\mu_p G_E^p/G_M^p$ at spacelike momentum transfer.
Secondly, both the strong near-threshold enhancement and the prominent oscillations in $|G_{\rm eff}|$
between the threshold at $t = 4m^2$ and $t\approx 6$~GeV$^2$ can be described after introducing
a certain number of broad poles above threshold in the spectral functions. 
These poles also generate the imaginary part
of the form factors in the physical region. Finally, we review the status of hyperon form factors and comment on the scale-dependence of the pion cloud.

We close with some open
questions related to baryon form factors that require more data and/or further investigations:
\begin{itemize}
	\item For the neutron data basis, a thorough analysis of the existing electron-deuteron
	and electron-$^3$He scattering data based on chiral effective field theory and
	including two-photon corrections should be performed. This would allow to consistently
	analyze the proton and neutron form factors based on the dispersive approach applied
	directly to cross section data.
	\item Data on $ep$ scattering or the polarization transfer at $Q^2 \gg 10\,$GeV$^2$ are
	urgently needed to investigate the onset of perturbative QCD. It will also be interesting
	to find out whether the form factor ratio really crosses zero as the present data seem
	to indicate.
	\item The precise experimental determination of the proton magnetic radius is
	urgently required to figure out whether a new "proton radius puzzle" exists.
	\item The current dispersion theoretical analysis of the nucleon electromagnetic form factor need to be expanded to incorporate the upcoming muon-proton scattering data from the MUSE~\cite{Downie:2014qna} and AMBER~\cite{Adams:2018pwt}, aiming to definitively resolve the proton radius.
    \item In the case of the hyperon form factors, the data basis in the timelike region is slowly increasing, mostly through measurements from the
    BESIII collaboration. Here, the interplay of final-state interactions in the proton-antiproton system and of genuine resonance contributions
    is an important ingredient, but at present the main unresolved  problems and challenges reside on the experimental side. In particular, more
    differential cross section data are urgently required.
\end{itemize}

\begin{ack}[Acknowledgments]%
The work of UGM and YHL was supported in
part by the Deutsche Forschungsgemeinschaft (DFG, German Research
Foundation) and the NSFC through the funds provided to the Sino-German Collaborative  
Research Center TRR~110 ``Symmetries and the Emergence of Structure in QCD''
(DFG Project-ID 196253076 - TRR 110, NSFC Grant No. 12070131001),
by the Chinese Aca\-de\-my of Sciences (CAS) through a President's
International Fellowship Initiative (PIFI) (Grant No. 2025PD0022),
and by the EU Horizon 2020 research and innovation programme,
STRONG-2020 project under grant agreement No. 824093. 
UGM was also supported by the ERC AdG EXOTIC (grant No. 101018170)
by the MKW NRW under the funding code No.~NW21-024-A.
HWH was supported by the
Deut\-sche Forschungsgemeinschaft (DFG, German
Research Foundation) -- Projektnummer 279384907 -- CRC 1245 and by the German Federal Ministry of Education and Research (BMBF) (Grants No. 05P21RDFNB and 05P24RDB).
\end{ack}


\bibliographystyle{Numbered-Style} 
\bibliography{reference}

\end{document}